\documentclass[preprint]{aastex62}

\usepackage{aas_macros}
\usepackage{subFigure}
\usepackage{graphicx}
\usepackage{natbib}
\usepackage{array}
\usepackage{xcolor}
\usepackage{amssymb}
\usepackage{xspace}

\newcommand\bsco{$\beta$ Sco\xspace}%
\newcommand\copr{{\it Copernicus}\xspace}%
\newcommand\fuse{{\it FUSE}\xspace}%
\newcommand\gara{$\gamma$ Ara\xspace}%

\newcommand\sigs{$\sigma_{\rm spec}$\xspace}%
\newcommand\sigf{$\sigma_{\rm fit}$\xspace}%
\newcommand\sigc{$\sigma_{\rm cont}$\xspace}%

\newcommand\jrot{\mbox{J$^{\prime \prime}$}\xspace}%
\newcommand\pr{$^{\prime}$\xspace}%
\newcommand\prpr{$^{\prime \prime}$\xspace}%
\newcommand\mh{H$_{\rm 2}$\xspace}%

\newcommand\Tk{T$_{01}$\xspace}
\newcommand\Te{T$_{\rm exc}$\xspace}

\newcommand\logt{log$_{\rm 10}$}%
\newcommand\logtm{{\rm log}_{\rm 10}}%

\newcommand\uderr[2]{$^{+#1}_{-#2}$} %

\newcommand\Tref[1]{Table~\ref{#1}\xspace}
\newcommand\Fref[1]{Figure~\ref{#1}\xspace}

\newcommand\pctc{$\Delta_{\%}$\xspace}

\begin{document}
\title{Revisiting the Temperature of the Diffuse ISM with CHESS Sounding Rocket Observations}

\author{Nicholas Kruczek}
\altaffiliation{nicholas.kruczek@colorado.edu}
\affiliation{Laboratory for Atmospheric and Space Physics, University of Colorado, 600 UCB, Boulder, CO 80309, USA}
\author{Kevin France}
\affiliation{Laboratory for Atmospheric and Space Physics, University of Colorado, 600 UCB, Boulder, CO 80309, USA}
\author{Keri Hoadley}
\affiliation{Cahill Center for Astrophysics, California Institute of Technology, 1216 E California Blvd, Pasadena, CA 91125, USA}
\author{Brian Fleming}
\affiliation{Laboratory for Atmospheric and Space Physics, University of Colorado, 600 UCB, Boulder, CO 80309, USA}
\author{Nicholas Nell}
\affiliation{Laboratory for Atmospheric and Space Physics, University of Colorado, 600 UCB, Boulder, CO 80309, USA}

\begin{abstract}
Measuring the temperature and abundance patterns of clouds in the interstellar medium (ISM) provides an observational basis for models of the physical conditions within the clouds, which play an important role in studies of star and planet formation. The Colorado High-resolution Echelle Stellar Spectrograph (CHESS) is a far ultraviolet rocket-borne instrument designed to study the atomic-to-molecular transitions within diffuse molecular and translucent cloud regions. The final two flights of the instrument observed $\beta^{1}$ Scorpii (\bsco) and $\gamma$ Arae. We present flight results of interstellar molecular hydrogen excitation on the sightlines, including measurements of the column densities and temperatures. These results are compared to previous values that were measured using the damping wings of low \jrot~\mh absorption features \citep{Savage77}. For \bsco, we find that the derived column density of the \jrot = 1 rotational level differs by a factor of 2--3 when compared to the previous observations. We discuss the discrepancies between the two measurements and show that the source of the difference is due to the opacity of higher rotational levels contributing to the \jrot = 1 absorption wing, increasing the inferred column density in the previous work. We extend this analysis to 9 \copr and 13 \fuse spectra to explore the interdependence of the column densities of different rotational levels and how the \mh kinetic temperature is influenced by these relationships. We find a revised average gas kinetic temperature of the diffuse molecular ISM of \Tk = 68 $\pm$ 13 K, 12\% lower than the value found previously.
\end{abstract}

\section{Introduction}
The raw materials for future star and planet formation reside in cool, dense clouds dispersed throughout the interstellar medium (ISM). These clouds span a comparably small temperature range and volume of interstellar space ($\sim$1--2\%), yet they contain approximately half of the mass of the ISM and are comprised of a variety of chemical constituents~\citep{Ferriere01}. The classification scheme proposed by~\cite{Snow06} distinguishes clouds using their local fraction of H and C in their atomic and molecular (\mh, CO) forms. Clouds that fall in the diffuse molecular to translucent region are rich in \mh and observations of the molecule provide insights into their chemical and physical conditions. These clouds are also optically thin enough that \mh can be observed in absorption in the far ultraviolet (FUV) bandpass. 

Past space-based and rocket-borne instruments (see, e.g. \citealt{Savage77,Rachford09,France13a}) have measured the column densities of \mh in these clouds across several rotational (\jrot) states.~\cite{Savage77} used the column densities of the \jrot = 0 (N(0)) and \jrot = 1 (N(1)) levels to calculate the kinetic gas temperatures (\Tk) of a sample of diffuse molecular sightlines. They found an average \Tk = 77 $\pm$ 17 K using a sample of 66 objects observed by the \copr satellite. This number has seen continued use as the point of comparison for more modern surveys of diffuse and translucent clouds~\citep{Rachford02,Burgh07,Sheffer08}.

We developed and launched a sounding rocket payload, called the Colorado High-resolution Echelle Stellar Spectrograph (CHESS). CHESS was designed to demonstrate state-of-the-art grating fabrication techniques and detector technologies to obtain high resolution spectra of ISM sightlines towards bright targets over a broad FUV bandpass (1000--1600 \AA). The science goals of CHESS were diverse but an error in the fabrication of one of its gratings resulted in decreased resolution, limiting the ability to quantify the carbon budget in interstellar clouds. Despite this issue, we were still able to obtain observations of \mh along three of the four sightlines observed during the lifetime of the instrument.

The first launch of CHESS (CHESS-1) was hampered by the low efficiency of an experimental echelle grating~\citep{Hoadley14}. This grating was replaced for CHESS-2, facilitating the observation of \mh along the line of sight towards $\epsilon$ Persei~\citep{Hoadley16,Hoadley19}. In this work we present the modeled column densities and excitation diagrams of \mh along the sightlines towards $\beta^1$ Scorpii (\bsco) and $\gamma$ Arae (\gara) that were observed on the final two launches of the CHESS sounding rocket, CHESS-3 and CHESS-4.

In \S\ref{overview} we provide a description of the CHESS payload, details on the targets and flights, and a summary of our modeling procedure. \S\ref{results} presents our modeled \mh column densities for \bsco and \gara and a comparison of our results to those of~\cite{Savage77}. The results of those comparisons prompt a re-examination of the conclusions drawn by Savage et al., which we detail in \S\ref{extended}. \S\ref{conclusions} provides a summary of our results.

\section{The CHESS Sounding Rocket} \label{overview}

\subsection{Targets}
The target for CHESS-3 was \bsco, a B0.5 V star at d = 161 pc with intermediate reddening (E(B-V) = 0.20, Av $\sim$ 0.6)~\citep{Savage77,Abt81}, indicating that the sightline may be sampling translucent material. H$_2$, C I, and CO were all detected by \copr along the line of sight to \bsco, however observations with higher sensitivity and spectral resolution were needed to understand the structure of the intervening matter~\citep{Federman80}. Additional studies found depletion of molecular material and ionized metal features, such as CO, Fe II, and Mg II, a result that is inconsistent with some nearby sightlines, such as $\zeta$ Oph and $\rho$ Oph~\citep{Bohlin83}. The fourth flight of CHESS observed \gara, a B1 I star at d = 689 pc that was chosen because it is known to display a variable and equatorially-enhanced stellar wind~\citep{Prinja97} that could potentially generate a population of rovibrationally excited H$_2$ at the wind/ISM interface.

\subsection{Instrument}
CHESS is an objective f/12.4 echelle spectrograph. The instrument design included the development of two novel grating technologies and the flight-testing of a cross-strip anode microchannel plate (MCP) detector~\citep{Beasley10}. The instrument was designed to achieve resolving powers $\ge$ 100,000 $\lambda$/$\Delta \lambda$ across a bandpass of 1000--1600 \AA~\citep{France16a}. The operating principle of CHESS is as follows:

\begin{itemize}

\item A mechanical collimator, consisting of an array of 10.74 mm $\times$ 10.74 mm $\times$ 1000 mm anodized aluminum tubes, provides CHESS with a total collecting area of 40 cm$^{2}$, a field of view (FOV) of 1.37$^{\circ}$, and allows on-axis stellar light through to the spectrograph.

\item A square echelle grating (ruled area: 102 mm $\times$ 102 mm), with a designed groove density of 87 grooves/mm and angle of incidence (AOI) of 63$^{\circ}$, intercepts and disperses the FUV stellar light into higher diffraction terms (m = 200--124). The grating is coated with aluminum and a lithium-flouride overcoat (Al+LiF).

\item Instead of using an off-axis parabolic cross disperser~\citep{Jenkins88}, CHESS employs a holographically-ruled cross dispersing grating with a toroidal surface figure. The cross disperser is ruled over a square area (100 mm $\times$ 100 mm) with a groove density of 351 grooves/mm and was designed to have a surface radius of curvature (RC) = 2500.25 mm and a rotation curvature ($\rho$) = 2467.96 mm. The grating spectrally disperses the echelle orders and corrects for grating aberrations~\citep{Thomas03}. The grating is coated with Al+LiF.

\item The cross-strip MCP detector~\citep{Vallerga10,Siegmund09} is circular in format, 40 mm in diameter, and capable of total global count rates of $\sim$10$^{6}$ counts/second. The cross-strip anode allows for high resolution imaging, with the location of a photoelectron cloud determined by the centroid of current readout from nine anode ``fingers" along the x and y axes.
\end{itemize}

\begin{figure}
\begin{center}
\includegraphics[width=0.95\textwidth]{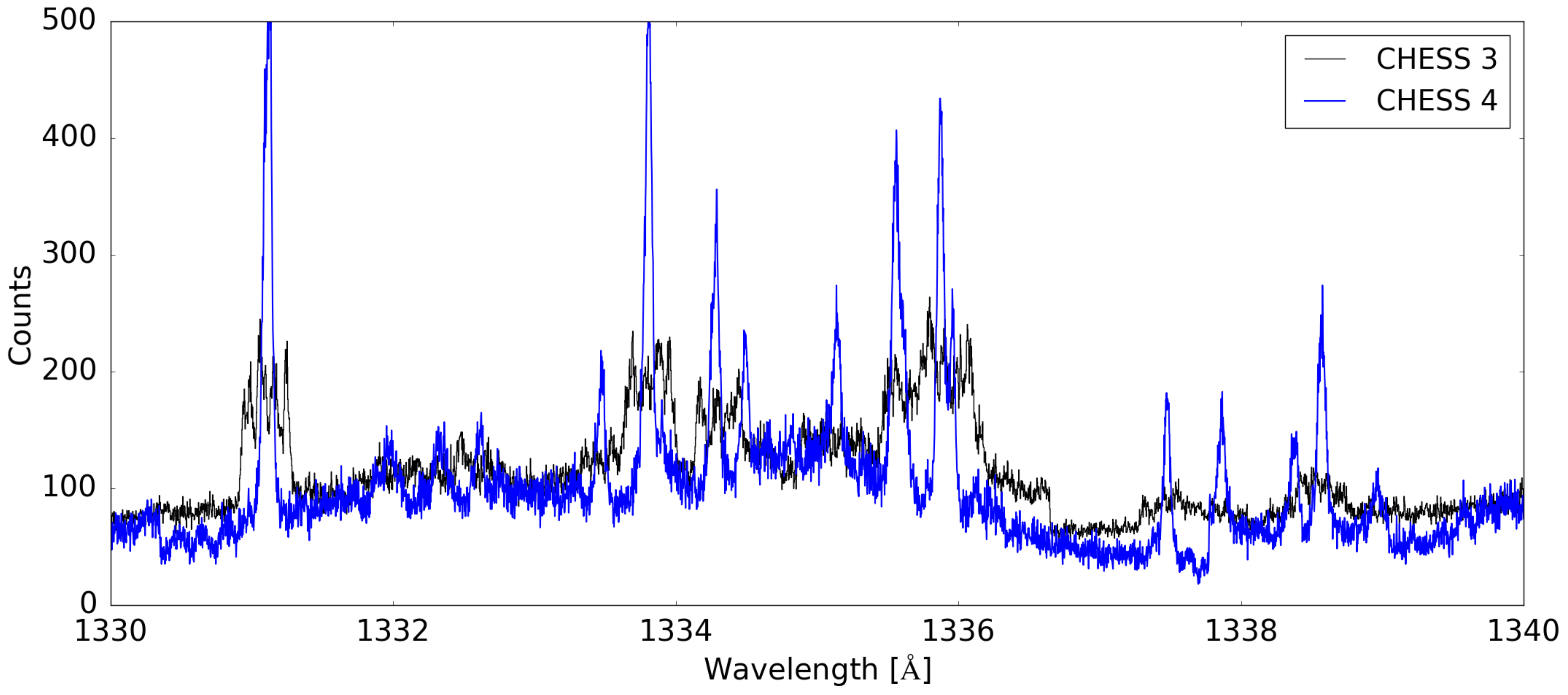}
\caption{A comparison of the CHESS-3 pre-flight spectrum (black) and the CHESS-4 pre-flight spectrum (blue).}
\label{labLSF}
\end{center}
\end{figure}

\begin{figure}
\centering
\subfigure{\includegraphics[height=0.48\textwidth]{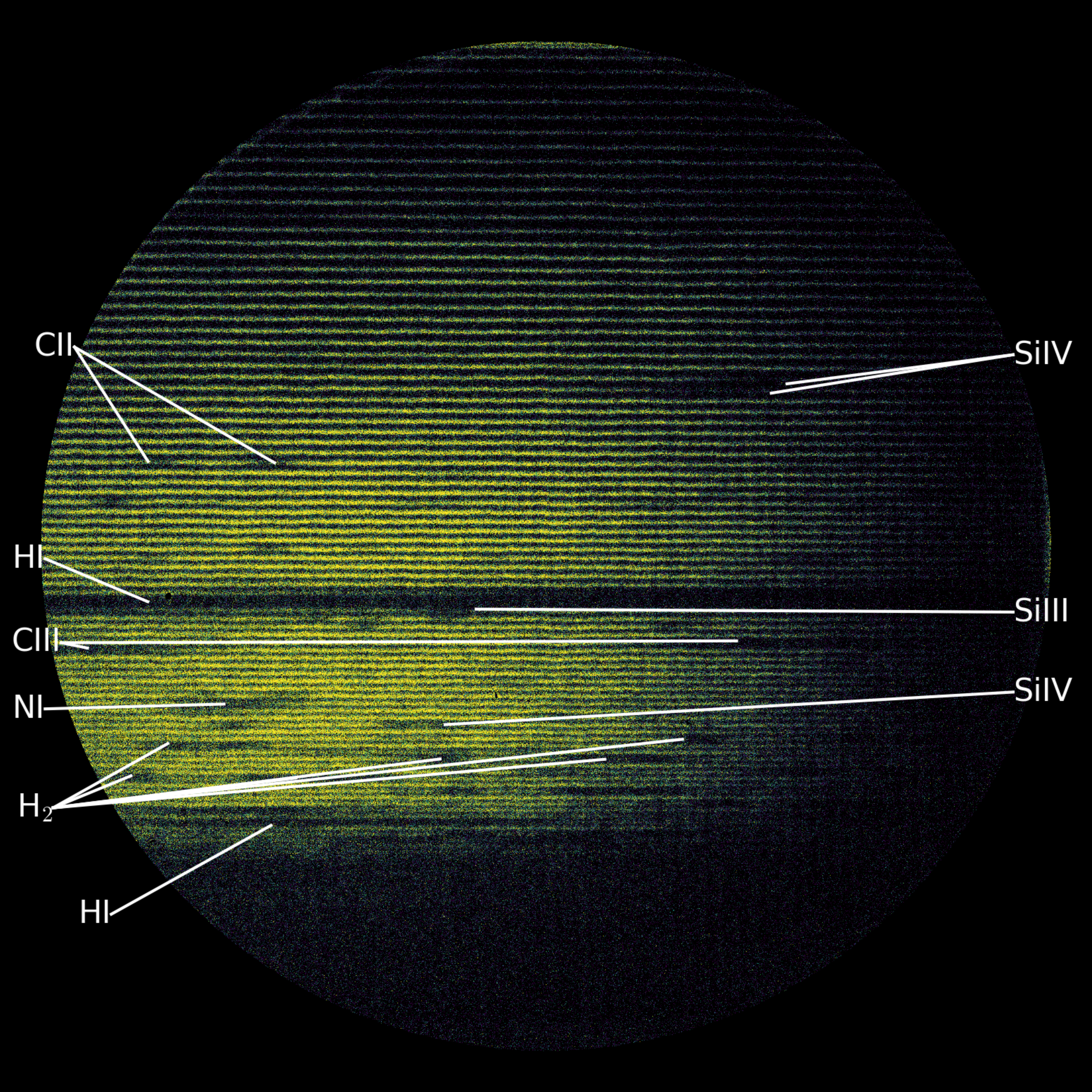}}
\subfigure{\includegraphics[height=0.48\textwidth]{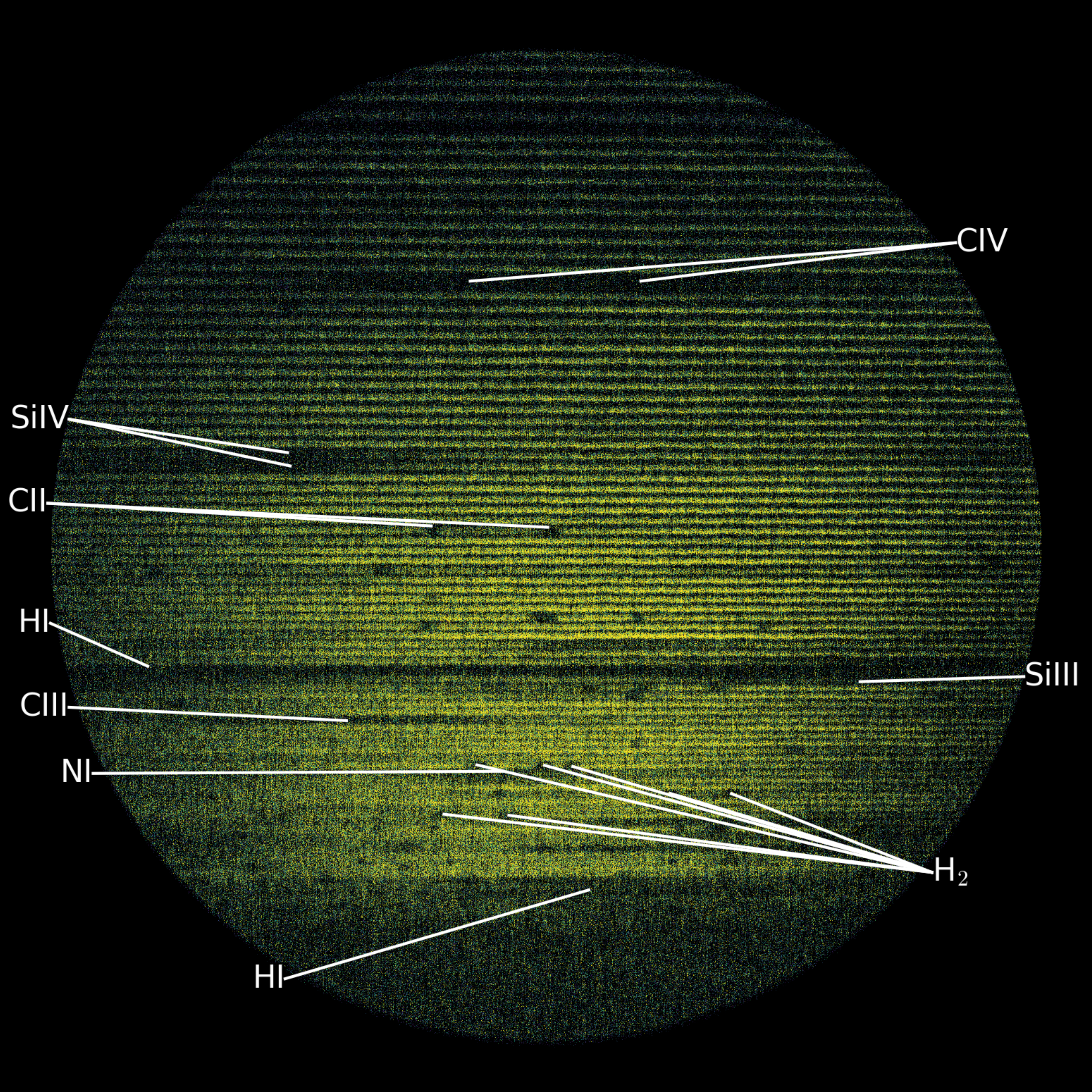}}
\caption{{\it Left:} The background subtracted flight spectrum of \bsco from CHESS-3. For this launch, we achieved $\sim$360 seconds of on target observing, with a peak count rate of 160,000 counts/second. After the on-target observations, we took a 30 second, off-target airglow measurement to use for background correction. {\it Right:} The background subtracted flight spectrum of \gara from CHESS-4. For this launch, we achieved $\sim$300 seconds of on-target observing, with a peak count rate of 130,000 counts/sec. We again ended our observations with a 30 second airglow measurement. Prominent stellar and ISM absorption features have been labeled in both images.}
\label{flightspecs}
\end{figure}

An error in the fabrication of the cross disperser resulted in the grooves being ruled 90$^{\circ}$ off, making it impossible to simultaneously minimize the widths of the spectral features and echelle orders. To avoid overlap in neighboring orders, we had to focus the instrument in a configuration that had a resolving power (R) well below the designed R $\sim$ 100,000. For the first three flights of CHESS, the maximum achievable R was $\sim$3,800. We mitigated this issue for CHESS-4 by mechanically shaping the surface of the echelle grating through the use of precisely torqued set screws. This change resulted in an increased R of $\sim$13,900. Our echelle bend procedure has been discussed previously~\citep{Kruczek18} and will be outlined in more detail in an upcoming publication~\citep{Kruczek19}. This resolution enhancement is demonstrated when comparing the 1D laboratory spectra (\Fref{labLSF}) and flight echellograms (\Fref{flightspecs}) of CHESS-3 and CHESS-4.

\subsection{Flight Details}
CHESS-3 was launched aboard NASA mission 36.323 UG from White Sands Missile Range (WSMR) on 26 June 2017 at 11:10pm MDT using a two-stage Terrier/Black Brant IX vehicle. The mission was deemed a comprehensive success. A single uplink maneuver was needed to properly align the star to the optical axis, meaning that the instrument was able to integrate for $\sim$360 seconds on-target, with an approximate count rate of 190,000 photons/sec. After the 360 second exposure, we moved to an off-target calibration position where we took a 30 second long exposure to obtain a measurement of the background Ly$\alpha$ and O I airglow that contaminated our on-target spectrum.

CHESS-4 was launched aboard NASA mission 36.333 UG from the Reagan Test Site on Roi-Namur in the Republic of the Marshall Islands on 17 April 2018 at 4:47 am MHT using a two-stage Terrier/Black Brant IX vehicle. The mission was deemed a comprehensive success. A single uplink maneuver was needed to initially align the star to the optical axis and we were able to integrate for $\sim$300 seconds on-target, with an approximate count rate of 125,000 photons/sec. We again moved to an off-target calibration position to obtain a $\sim$40 second long background exposure.

\subsection{Analysis}
The construction of the 1D continuum-normalized spectra from the flight echellograms has been discussed in detail previously and will not be repeated here~\citep{Hoadley14,Hoadley16,Kruczek17,Kruczek18}. Modeling of the \mh absorption features within these spectra was done using the {\it H2ools} optical depth templates~\citep{McCandliss03}. These templates are calculated for integer values of $b$ = 2--20 km s$^{-1}$, \jrot = 0--15, v\pr = 0--18 (for the Lyman band), and v\prpr = 0--3 and they are useful for N(\mh) $\lesssim$ 10$^{21}$ cm$^{2}$. We expect $b >$ 12 km s$^{-1}$ to be unphysically large for the sightlines analyzed in this work so, to minimize the possible parameter space, we set an upper limit on $b$ at that value. To allow for non-integer $b$ values, we performed a weighted average of the templates above and below the non-interger value, where the weight was determined using 1 - $\mid$ $b_{\rm int}$ - $b_{\rm non-int}$ $\mid$.

The column densities for \jrot $>$ 7 were expected to be smaller than the uncertainities in the observations and so we only modeled, at most, up to N(7). This decision is supported by our measured uncertainites in even the \jrot = 6 and 7 column densities (see, e.g. \Fref{bsco_excit} in \S\ref{bsco_res}). To avoid contamination from the Werner transitions, we restrict our fitting routine to the (0-0) to (4-0) Lyman bands. This restriction is imposed by limiting the bandpass to $\lambda \lambda$ 1046--1120 \AA, in general.

Our code accepted an initial guess for N(0--7) and $b$ and, using the NumPy leastsqs routine, performed a least squares minimization between the observed spectrum and a convolution of the {\it H2ools} templates with our expected instrument profile. For CHESS-3, we used an R = 3,800 Guassian for the instrument profile. For CHESS-4 this was updated to R = 13,900. We masked known stellar and ISM absorption features before performing the analysis, using~\cite{Pellerin02} as a guide to identify the lines. The resulting N(\jrot) values were fed back into the template and instrument convolution to generate a model spectrum. The column densities were further used to calculate the temperatures of the \jrot states. N(0) and N(1) are used, along with the Boltzmann equation, to calculate \Tk. The excitation temperature (\Te) of the higher (\jrot $>$ 2) states is calculated again assuming the levels follow a Boltzmann distribution. In this case, the temperature is found using the slope of a linear fit to ln(N(\jrot)/g$_{\rm \jrot}$), where g$_{\rm \jrot}$ is the degeneracy of the \mbox{J$^{\prime \prime}$}$^{th}$ level.

There are several different error contributions to the column density determination that, for clarity, we will name separately. The first error is the photometric uncertainty, which we will refer to as \sigs. The fitting routine, which accounts for \sigs, returns uncertainties in the resulting modeled values. We will refer to this error as \sigf. Finally, there is an error associated with our continuum placement. While we try to construct the continuum by bisecting the flux measurements in unabsorbed portions of the spectrum the physical continuum level could be within $\pm$ 1\sigs. To quantify the effect of this uncertainty, we repeat the fitting procedure on the spectra that are produced when the continuum is moved $\pm$ 1\sigs. The average 1\sigs level was determined by calculating the standard deviation of a representative unabsorbed portion of the spectrum, and was found to be $\sim$0.1 in normalized flux units. \sigc is then equal to the differences in the measured column densities and $b$ values found at the raised and lowered continuum positions. \sigc is a conservative estimate of the error in $b$ and N because it maximizes the uncertainty in the continuum. It is statistically unlikely (ignoring unidentified systematics) that our continuum placement was 1\sigs high/low across the entire bandpass and, therefore, the values produced in those limits gives the largest possible change in $b$ and N. In practice, we indeed found that \sigc was the largest of the measured uncertainties and so we use those values as our quoted errors.

Two errors were calculated for \Tk, one was found using the difference between the modeled temperature and the temperature calculated using the $\pm$ 1\sigc results. The second was found using the difference between the modeled temperature and the temperature calculated using N(0) $\pm$ 1\sigf and N(1) $\mp$ 1\sigf. The final error was chosen to be the maximum value between the two methods. Our \Te measurements are less well constrained and were determined using only the difference found using the $\pm$ 1\sigc results.

\section{CHESS-3 and CHESS-4 Flight Results} \label{results}
\subsection{\bsco}
\subsubsection{CHESS result} \label{bsco_res}
\begin{figure}
\begin{center}
\includegraphics[width=0.98\textwidth]{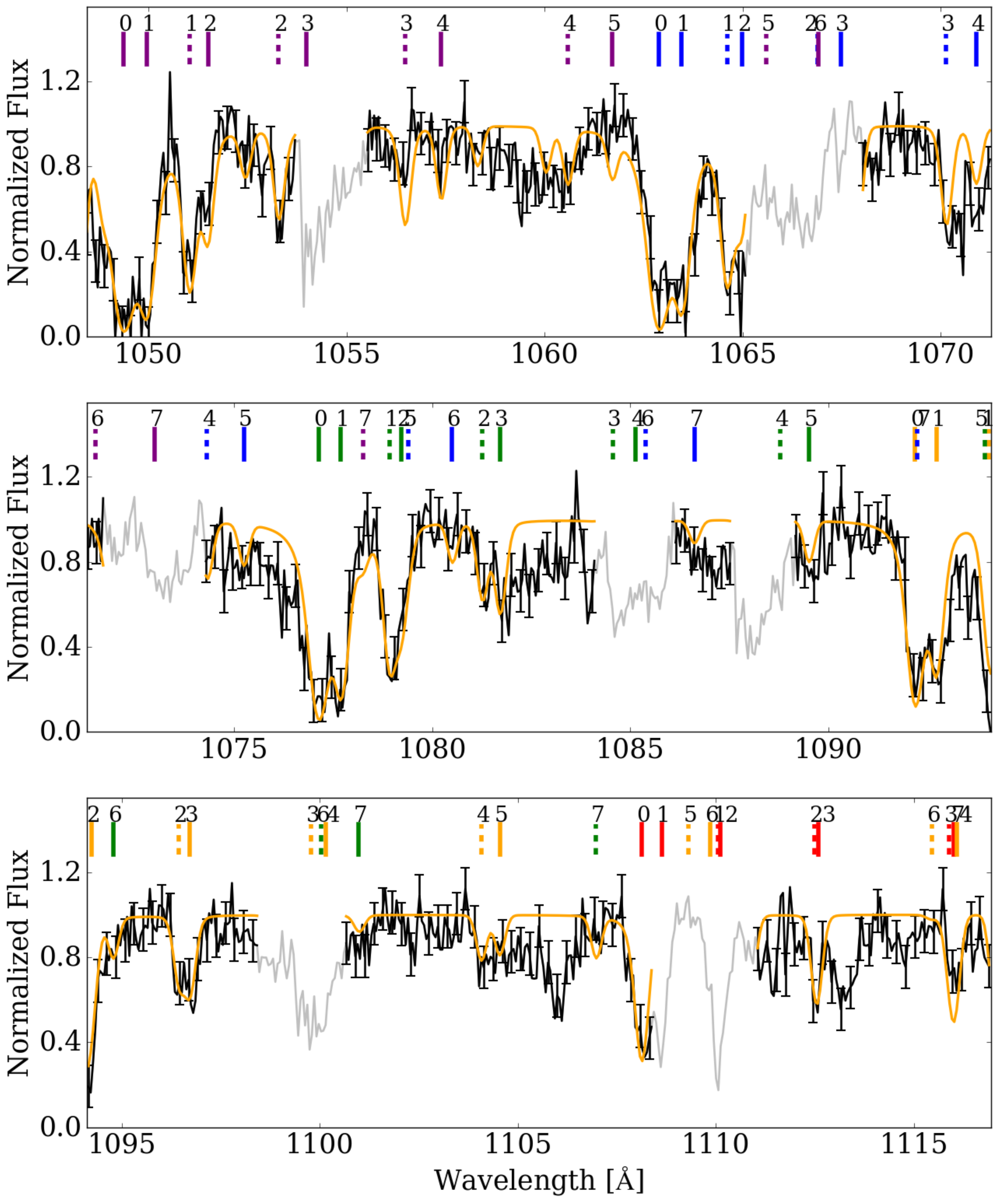}
\caption{The extracted 1D flight spectrum for \bsco, plotted in black, with our model overplotted in orange. Regions of the spectrum that were masked are plotted in gray. The vertical ticks indicate the positions of the \mh absorption features, up to \jrot = 7, from the following vibration bands: v$^{\prime}$-v$^{\prime \prime}$ = 0-0 (red), 1-0 (orange), 2-0 (green), 3-0 (blue), and 4-0 (purple). The number above each tick display the \jrot level, with a solid tick indicating a R(\jrot) transition and a dashed tick indicating a P(\jrot) transition.}
\label{bsco_chess}
\end{center}
\end{figure}

Figure~\ref{bsco_chess} shows the continuum normalized flight spectrum of \bsco over the bandpass of interest for H$_2$ absorption features. Overplotted in orange is the model absorption profile that was found by fitting the H$_2$ features. The final spectrum has been binned down to d$\lambda$ $\sim$ 0.06 \AA~per bin, which is about 5 bins per resolution element. A summary of the fit parameters are listed in \Tref{bsco_fitres}. We confirmed the robustness of our \sigc assumption for these results by identifying features in \Fref{bsco_chess} that appear by eye to disagree significantly with the model and then masking them before remodeling the column densities. For the four lines tested -- (4-0) P(3), (4-0) R(4), (4-0) P(4), and (1-0) R(2) -- the column densities found using the masked spectrum remained within the quoted \sigc errors.

\begin{figure}
\begin{center}
\includegraphics[width=0.6\textwidth]{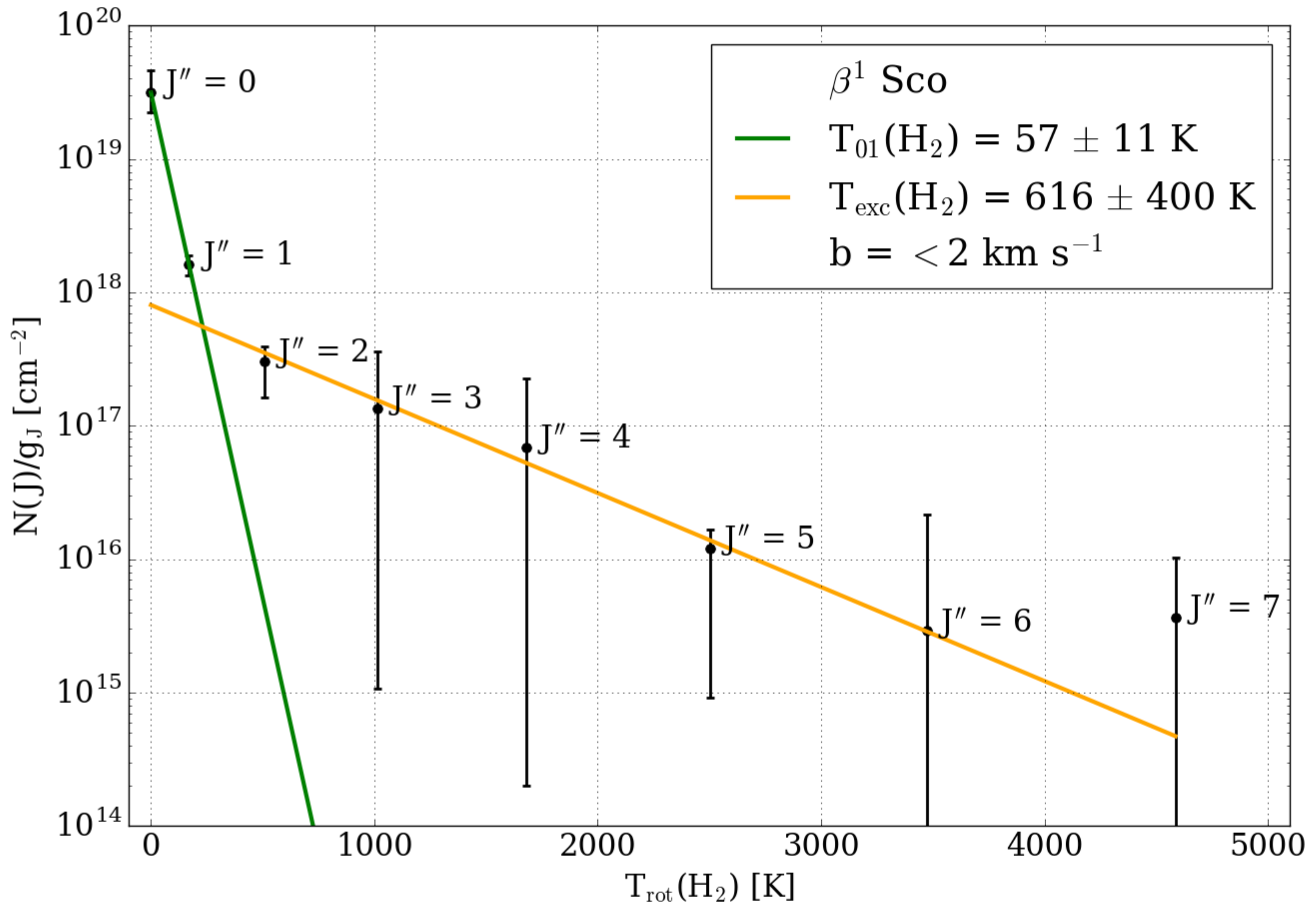}
\caption{\mh excitation diagram derived from our spectral profile fitting of \bsco using rotational levels \jrot = 0--7. Our calculated \Tk and \Te are listed in the legend. Their corresponding lines are plotted in green (for \Tk) and orange (for \Te).}
\label{bsco_excit}
\end{center}
\end{figure}

Figure~\ref{bsco_excit} shows the excitation diagram for our modeled spectrum, where we find \Tk = 57 $\pm$ 11 K and \Te = 607 $\pm$ 400 K. Additional uncertainties on N(\jrot $\gtrsim$ 3) and \Te that are not reflected in our error calculation arise due to the limitations placed on $b$ in our fitting routine. The first issue comes from the boundaries imposed by the {\it H2ools} templates, which span $b$ = 2--12 km s$^{-1}$. The $b$ value produced by our fitting routine equaled 2 km s$^{-1}$ indicating that the true $b$ value could be lower. This would impact the higher \jrot column densities that lie close to or on the flat portion of the curve of growth, since a lower $b$ favors a larger N(\jrot). The second issue is that our model assumes a single $b$ value for all \jrot states. Previous observations have shown evidence of an increasing $b$ with \jrot~\citep{Spitzer73,Jenkins97,Jenkins00,Lacour05}. Therefore, a systematic error in our N(\jrot) measurements could exist for the larger \jrot levels.

\begin{table}[h]
\centering
\caption{CHESS-3 \bsco Fit Results
\label{bsco_fitres}}
\begin{tabular}{ c  c  c  c  c  c }
\hline
\rule{0pt}{2.5ex} b & \logt N(\mh) & \logt N(0) & \logt N(1) & \logt N(2) & \logt N(3) \\
\lbrack km s$^{-1}$] & [\logt~cm$^{-2}$] & [\logt~cm$^{-2}$] & [\logt~cm$^{-2}$] & [\logt~cm$^{-2}$] & [\logt~cm$^{-2}$] \\ \hline
\rule{0pt}{3ex} $\leq$2 & 19.71\uderr{0.17}{0.16} & 19.50\uderr{0.17}{0.15} & 19.17\uderr{0.07}{0.08} & 18.12\uderr{0.11}{0.27} & 18.37\uderr{0.43}{2.10} \\ \hline
\\ \hline
\rule{0pt}{3ex} \logt N(4) & \logt N(5) & \logt N(6) & \logt N(7) & \Tk & \Te \\
\lbrack \logt~cm$^{-2}$] & [\logt~cm$^{-2}$] & [\logt~cm$^{-2}$] & [\logt~cm$^{-2}$] & [K] & [K] \\ \hline
\rule{0pt}{2.5ex} 17.81\uderr{0.52}{2.53} & 17.60\uderr{0.14}{1.12} & 16.78\uderr{0.87}{6.78} & 17.19\uderr{0.45}{7.19} & 57 $\pm$ 11 & 607 $\pm$ 400 \\ \hline
\end{tabular}
\end{table}

\subsubsection{Comparison to~\cite{Savage77}} \label{bsco_comp}
~\cite{Savage77} (S77; hereafter) used the U1 channel of the \copr satellite to measure N(0) and N(1) along the line of sight to \bsco. After correcting for wavelength offsets and scatter, they derived the column densities by dividing the observed spectrum by predicted line shapes that were functions of column density. The best fit was determined by the column densities that best canceled out the absorption features. They measured \logt N(0) = 19.46 and \logt N(1) = 19.58 with a log error of $\pm$0.06 for each value. This result agrees well with our CHESS analysis for N(0) but our N(1) values disagree by $\sim$0.4 dex.

Apart from the host of uncertainties inherent to both data sets, a potential source for this discrepancy comes from the fact that S77 only used three lines: (1-0)R(0), (1-0)R(1), and (1-0)P(1), to make their measurement. While they state that one must acknowledge the existence of the (1-0)R(2) line, which partially overlaps with P(1), it is unclear what, if any, measures were taken to account for it. Therefore, their results may have favored larger N(1) values, since a larger column density would better account for some of the absorption produced by the R(2) line. To explore this further, we reproduce the analysis of S77 but use the CHESS fitting routine to model the column densities and compare the results when R(2) is and is not included.

We follow the procedure described by S77 to produce the \bsco spectrum, using their published plot as a guide (Figure 2 in S77). After correcting for background levels and wavelength offsets, we created a continuum using a linear fit between the regions of peak counts on either side of the absorption features. The resulting continuum normalized spectrum was fit using the CHESS analysis code. The instrument profile used in the convolution was a Gaussian with a full width at half-maximum of 0.051 \AA~\citep{Drake76}. S77 mentions using a ``flat-top Gaussian'', but insufficient detail is provided on the width of the flat portion. Given the stellar source profile shown in Figure 1 of~\cite{Drake76}, this flat portion is smaller than the pixel d$\lambda$ of the data set, so we do not expect excluding it will significantly impacts our results.

\begin{figure}
\begin{center}
\includegraphics[width=0.6\textwidth]{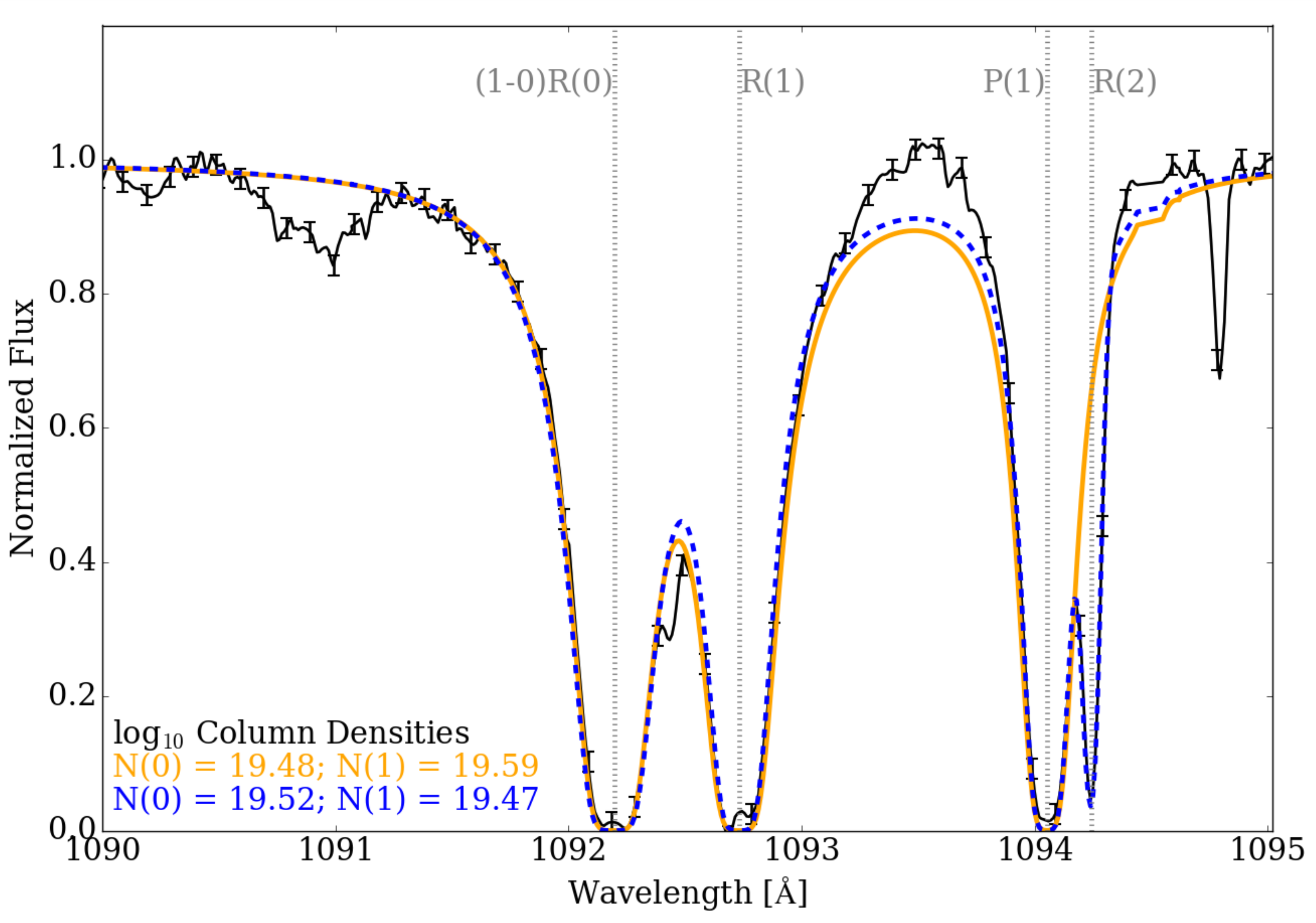}
\caption{The model results for the \bsco~\copr data using the CHESS analysis routine. The model produced using only the R(0), R(1), and P(1) lines is shown in orange. The model when R(2) is also included is plotted as a blue dashed line. The resulting column densities are shown in the lower left.}
\label{my_bsco}
\end{center}
\end{figure}

Figure~\ref{my_bsco} shows two resulting models of the \bsco~\copr spectrum, one where we fit the same lines as S77 and another where we include the (1-0)R(2) line. In both cases, the models do not agree well with the continuum region from 1093.1--1093.9 \AA. This is in the center of the stray light region of the \copr instrument~\citep{Rogerson73}, so we suspect that we are not fully accounting for the background throughout it. The purpose of this exercise was to recreate the S77 results, not ensure that we are properly measuring the column densities within it. For that reason, no effort was made to correct for the levels in that region. Table~\ref{bsco_summ} provides a summary of the various \bsco measurements performed in this work as well as from S77. We see that our initial recreation of the \copr measurement agrees well with their result, without accounting for the R(2) line in any capacity. We also observe the predicted decrease (by 0.12 dex; $\sim$30\%) in \logt N(1) once the R(2) line is included in the analysis. \logt N(0) also increases by 0.04 dex, which is a consequence of the overlap in the R(0) and R(1) lines. 

\begin{table}
\caption{A Summary of \bsco \mh Analyses
\label{bsco_summ}}
\centering
\begin{tabular}{ c  c  c  c  c  c }
\hline
\rule{0pt}{2.5ex} Source & \logt N(\mh) & \logt N(0) & \logt N(1) & \logt N(2) & T$_{01}$ \\
		 & [\logt~cm$^{-2}$] & [\logt~cm$^{-2}$] & [\logt~cm$^{-2}$] & [\logt~cm$^{-2}$] & [K] \\ \hline
\rule{0pt}{3ex} Savage 77 & 19.83 & 19.46 $\pm$ 0.06 & 19.58 $\pm$ 0.06 & -- & 88 \\[1.1ex]
{\it Copr.} no R(2)$^{\rm a}$ & 19.84\uderr{0.04}{0.04} & 19.48\uderr{0.04}{0.03} & 19.59\uderr{0.03}{0.05} &  -- & 87 $\pm$ 8 \\[1.1ex]
{\it Copr.} w/ R(2)$^{\rm a}$ & 19.79\uderr{0.05}{0.04} & 19.52\uderr{0.03}{0.04} & 19.47\uderr{0.04}{0.05} & 16.43\uderr{0.89}{0.31} & 74 $\pm$ 6\\[1.1ex]
CHESS-3 to \jrot = 1 & 19.80\uderr{0.22}{0.19} & 19.58\uderr{0.26}{0.21} & 19.39\uderr{0.14}{0.16} & -- & 64 $\pm$ 6\\[1.1ex]
CHESS-3 to \jrot = 2 & 19.77\uderr{0.22}{0.18} & 19.62\uderr{0.26}{0.22} & 19.17\uderr{0.03}{0.08} & 18.44\uderr{0.32}{0.42} & 52 $\pm$ 7 \\[1.1ex]
CHESS-3 to \jrot = 7 & 19.71\uderr{0.17}{0.16} & 19.50\uderr{0.17}{0.15} & 19.17\uderr{0.07}{0.08} & 18.12\uderr{0.11}{0.27} & 57 $\pm$ 11 \\ \hline
\multicolumn{6}{l}{{\bf Note:} Column density errors were calculated using \sigc.} \\
\multicolumn{6}{l}{$^{\rm a}$ Fits to the \copr data using the CHESS analysis code.} \\
\end{tabular}
\end{table}

We quantify the magnitude of this change by measuring the percent change (\pctc) in N(0) and N(1) between the fit with \jrot = 2 and without. We define \pctc as:
\begin{equation}
\Delta_{\%}{\rm N}(\jrot) = 100 \times \frac{\mbox{N(\jrot)}_{x} - \mbox{N(\jrot)}_{x-1}}{\mbox{N(\jrot)}_{x}}
\label{dpcnt}
\end{equation}
Where N(\jrot)$_{x}$ is the column density of rotational level \jrot when $x$ levels are included in the fit and N(\jrot)$_{x-1}$ is the same but for the case that $x$-1 levels are included. Using this equation and our measured column densities from the \copr data, we find \pctc{}N(0) = 8.0 $\pm$ 3.6\% and \pctc{}N(1) = -32.3 $\pm$ 2.5\%.

The errors on \pctc are calculated through standard propagation of error techniques but in this case we use \sigf as the error for the column densities, not \sigc. This is because \sigc captures the error in our continuum placement, which would have a similar impact on the column density measurements with and without \jrot = 2 and so the error would be degenerate between the two measurements. $\mbox{N(\jrot)}_{x}$ and $\mbox{N(\jrot)}_{x-1}$ result from fits to identical data, the only difference being the number of \jrot levels included their models. This means that there also exists a degeneracy in \sigf between the two models. As we have seen, the column density for a given \jrot will rely on the column densities of the other \jrot levels that are considered, which complicates attempts to disentangle the error degeneracy between the two models. To avoid overconstraining our resulting errors by making assumptions about the amount of degeneracy, we choose to adopt the \sigf values and treat them as a worst-case estimate of the error.

Even with this additional correction to the S77 N(1) value, our measurements still disagree. The difference in \logt N(2) of almost 2 dex indicates that line blending between P(1) and R(2) could still be occurring. This is supported by the good agreement between our N(\mh) and N(0) values as well as the fact that the excitation diagram for a low N(2) value, on the order of the 10$^{16.50}$ cm$^{-2}$ that we measured using the \copr data, would look nonphysical when compared to the values of N(0) and N(1).

\subsection{\gara}
\subsubsection{CHESS results}
\begin{figure}
\begin{center}
\includegraphics[width=\textwidth]{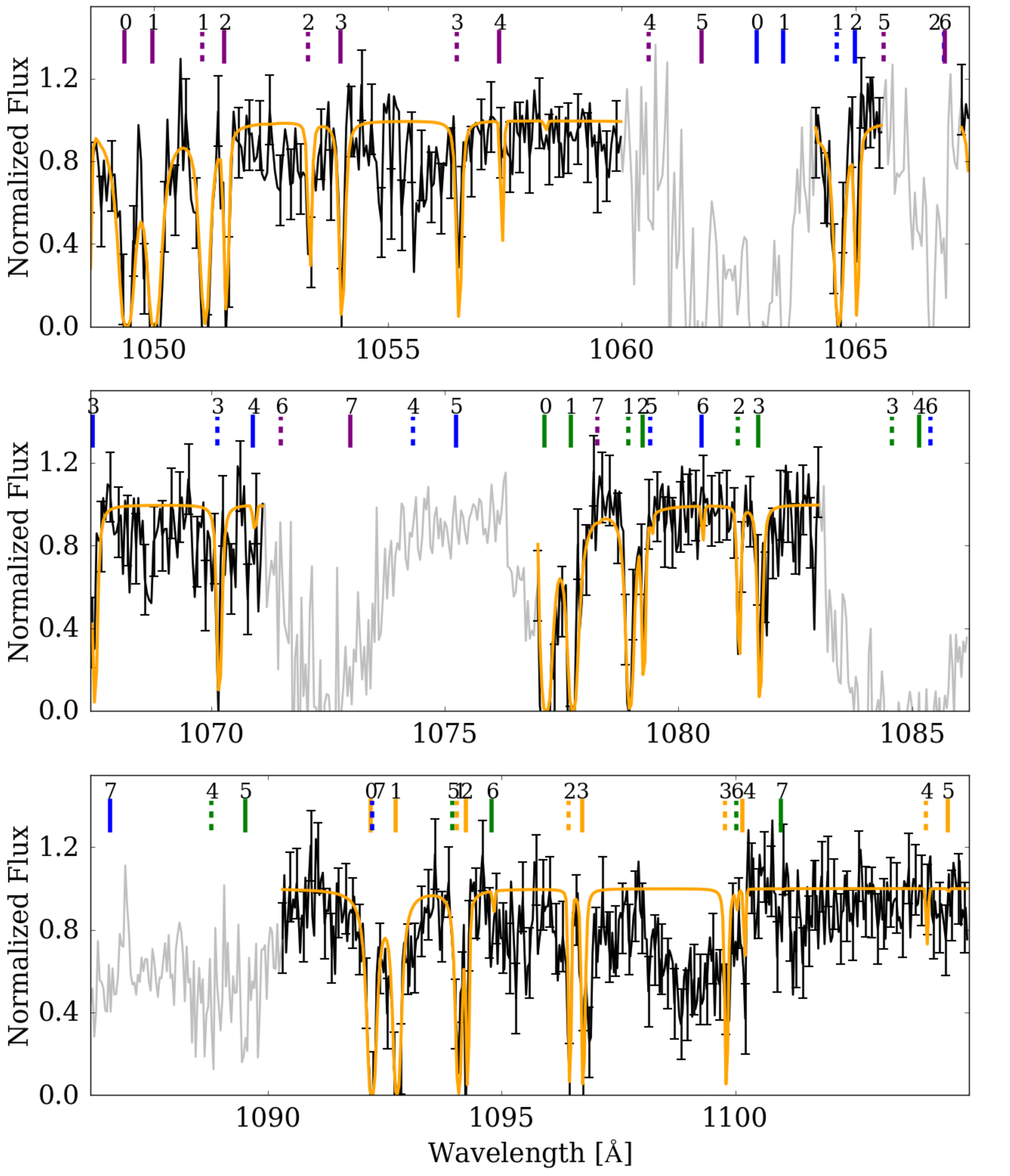}
\caption{The extracted 1D flight spectrum for \gara, plotted in black, with our model overplotted in orange. Regions of the spectrum that were masked are plotted in gray. The vertical ticks indicate the positions of the \mh absorption features, up to \jrot = 7, from the following vibration bands: v$^{\prime}$-v$^{\prime \prime}$ = 0-0 (red), 1-0 (orange), 2-0 (green), 3-0 (blue), and 4-0 (purple). The number above each tick display the \jrot level, with a solid tick indicating a R(\jrot) transition and a dashed tick indicating a P(\jrot) transition. }
\label{gara_chess}
\end{center}
\end{figure}

\begin{figure}
\begin{center}
\includegraphics[width=0.6\textwidth]{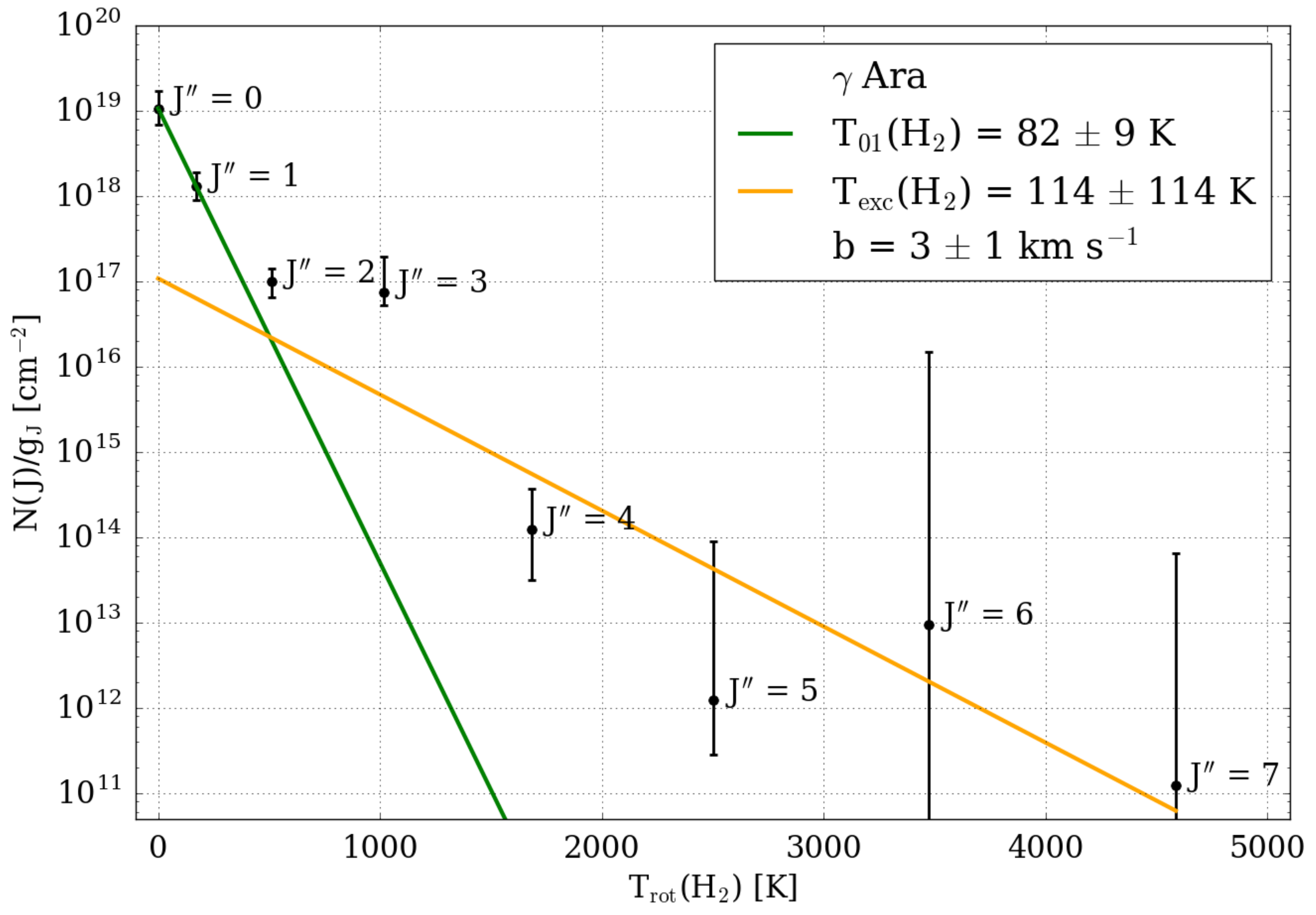}
\caption{\mh excitation diagram derived from our spectral profile fitting of \gara using rotational levels \jrot = 0--7. Our calculated \Tk and \Te are listed in the legend. Their corresponding lines are plotted in green (for \Tk) and orange (for \Te).}
\label{gara_excit}
\end{center}
\end{figure}

\Fref{gara_chess} shows the continuum normalized flight spectrum of \gara. Overplotted in orange is the modeled \mh absorption profile. The spectrum was produced using the same methods described in \S\ref{bsco_res}. In this case it has been binned down to d$\lambda$ $\sim$ 0.04 \AA~per bin, which is about 2 bins per resolution element. The resulting fit parameters are listed in \Tref{gara_fitres}. Figure~\ref{gara_excit} shows the excitation diagram for our modeled spectrum. We tested the validity of our \sigc assumption using the same process and features identified in \S\ref{bsco_res} and again found that the masked spectra produced column densities that were within the quoted \sigc errors.

The spectrum was heavily impacted by stellar and ISM absorption features. In this case, the effect was larger than it was for \bsco due to the wind-broadening of the stellar lines and \gara's larger distance. Attempts to fit these feature along with the continuum did not improve the resulting \mh fits, so we instead opted to mask them. This included S IV 1062, 1072, and 1073 \AA, Ar I 1066 \AA, and the N II complexes around 1084 and 1085 \AA. The Si III triplet near 1110 \AA~obscured the low \jrot (0-0) vibrational band features and so we exclude wavelengths longer than 1105 \AA. There is an additional unidentified absorption feature on the short wavelength side of (2-0)R(0), near 1076 \AA, that we observe in our data as well as the \copr data. This feature is masked in both analyses.

Due to a combination of factors (the star is more distant, our pointing drifted during flight, and we had higher resolution), the S/N of the \gara data is lower than that of \bsco and this is reflected in the resulting fits. In particular, our modeled N(3) seems relatively high compared to N(2), and that  larger value could be influencing the smaller modeled column densities at higher \jrot states. As such, we feel confident comparing our modeled N(0) and N(1) values to the results of S77, but we do not expect that our higher \jrot column densities or \Te measurement provide a meaningful constraint to the \gara sightline.

\begin{table}
\caption{CHESS-4 \gara Fit Results
\label{gara_fitres}}
\centering
\begin{tabular}{ c  c  c  c  c  c }
\hline
\rule{0pt}{2.5ex} b & \logt N(\mh) & \logt N(0) & \logt N(1) & \logt N(2) & \logt N(3) \\
\lbrack km s$^{-1}$] & [\logt~cm$^{-2}$] & [\logt~cm$^{-2}$] & [\logt~cm$^{-2}$] & [\logt~cm$^{-2}$] & [\logt~cm$^{-2}$] \\ \hline
\rule{0pt}{3ex} 3.0 $\pm$ 1.0 & 19.39\uderr{0.20}{0.19} & 19.03\uderr{0.16}{0.17} & 19.07\uderr{0.15}{0.18} & 17.69\uderr{0.41}{0.16} & 18.20\uderr{0.48}{0.59} \\ \hline
\\ \hline
\rule{0pt}{2.5ex} \logt N(4) & \logt N(5) & \logt N(6) & \logt N(7) & \Tk & \Te \\
\lbrack \logt~cm$^{-2}$] & [\logt~cm$^{-2}$] & [\logt~cm$^{-2}$] & [\logt~cm$^{-2}$] & [K] & [K] \\ \hline
\rule{0pt}{3ex} 15.04\uderr{1.87}{0.63} & 13.61\uderr{3.20}{12.53} & 14.09\uderr{2.71}{14.09} & 12.77\uderr{3.64}{12.77}  & 82 $\pm$ 9 & 114 $\pm$ 114 \\ \hline
\end{tabular}
\end{table}

\subsubsection{Comparison to~\cite{Savage77}} \label{gara_savage}

S77 also used the \copr satellite to measure the \mh column densities along the line of sight to \gara. In this case, the U2 channel was used. This channel has lower resolution but the scan it produced is complete over a bandpass $\lambda \lambda$ 1040--1120 \AA. They analyze the same vibrational bands, (0-0) to (4-0), that we covered in the CHESS analysis. They treated each band separately, obtaining a modeled N(0) and N(1) in each case using the R(0), R(1), and P(1) features. They then averaged their results to obtain values of \logt N(0) = 18.93 $\pm$ 0.23 and \logt N(1) = 18.94 $\pm$ 0.23. Our CHESS results and those of S77 are in better agreement in this case, differing by about 0.1 dex ($\sim$20\%), which is well within the error bars of both measurements. Nonetheless, that agreement is assuming that the inclusion of R(2) would not impact the S77 result, which was not the case for \bsco and disagrees with expectations. To test this, we again attempt to recreate the analysis of S77.

Unlike the \bsco case, S77 did not publish a plot of their final background corrected \gara spectrum and so we lack a similar point of comparison when attempting to recreate their results, instead relying on their written procedure. The instrument profile in this case was described as a trapezoid with FWHM = 0.2 \AA~\citep{Savage77}. No additional detail on the shape of the trapezoid was given. Instead, we found that the origin and shape of the U1 profile resulted from the convolution of the instrument entrance and exit slits~\citep{Jenkins75,Drake76}. Following that, we convolved two box car functions that were the width of the entrance and exit slits of the U2 channel (24 and 96 $\mu$m, respectively), and then scaled the resulting function so that it had a FWHM of 0.2 \AA. There are additional corrections that should be made to the perfect trapezoidal shape due to effects such as the slight variation in the diffraction angle for different wavelengths off the grating. This results in light entering the slit at slightly different angles. The magnitude of these effects are assumed to be small (on the order of 3\% of the total width in the U1 case) and, given that the d$\lambda$ = 0.1 \AA~of the U2 data is comparably large, we choose to ignore any influence from them. The continuum was again created using a linear fit across each individual absorption complex.

\begin{figure}
\begin{center}
\includegraphics[width=\textwidth]{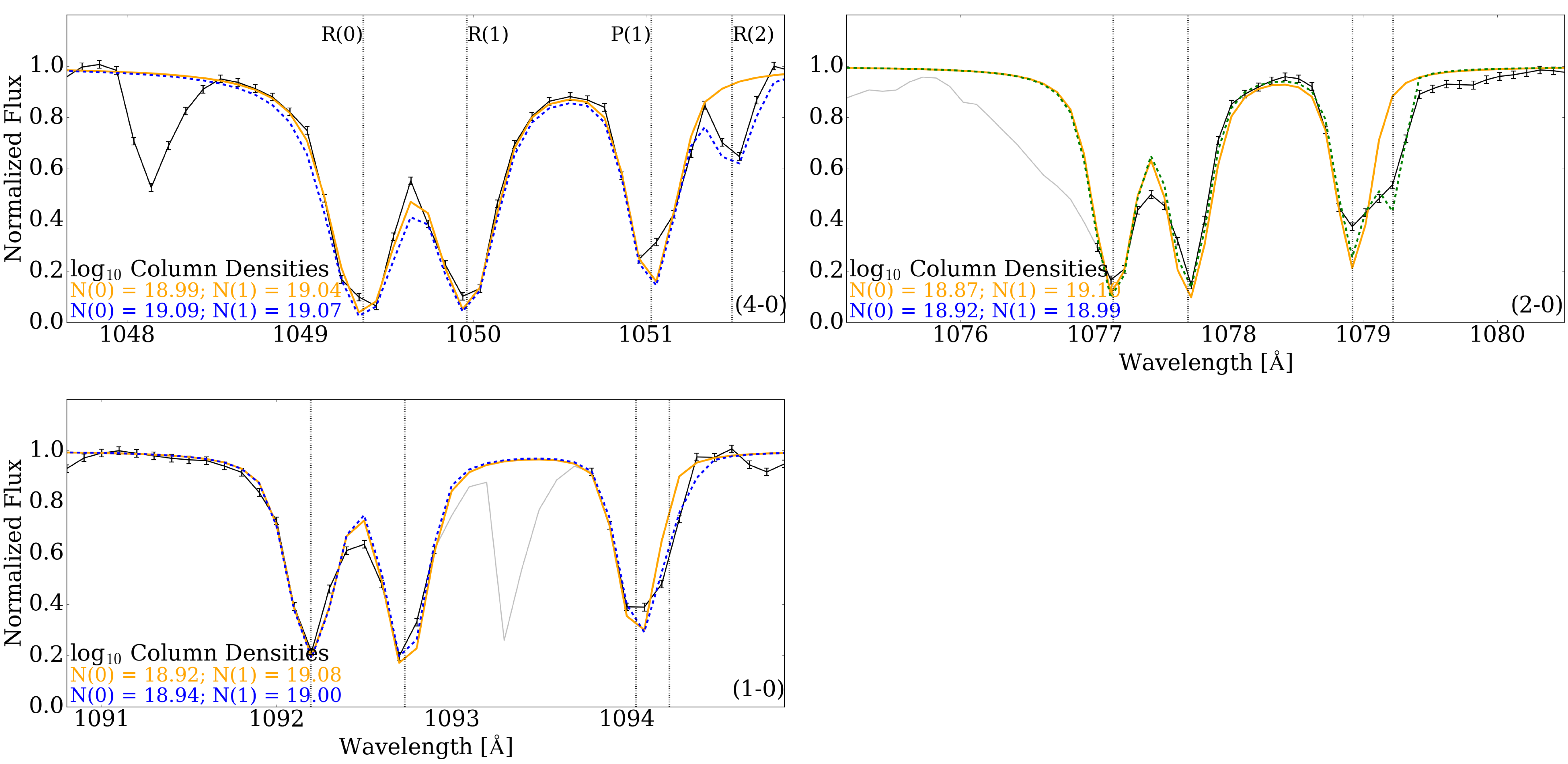}
\caption{Fits to the \gara~\copr data using the CHESS flight data analysis code. The spectrum is plotted in black, with regions that were excluded from the analysis plotted in gray. Each figure shows the fits for a single vibrational band, indicated by the designation in the lower right. The orange line shows the fit when the R(2) line is excluded and the blue line shows when it is included. The column densities are listed in the lower left of each figure.}
\label{my_gara}
\end{center}
\end{figure}

\Fref{my_gara} shows the resulting fits for the three vibrational bands studied in the \gara~\copr spectrum. The (0-0) and (3-0) bands were excluded due to contamination from non-\mh absorption features. Like we saw in the CHESS spectrum, the (2-0)R(0) line has excess absorption on the blue end of the feature. As was done in \S\ref{bsco_comp}, we performed one fit using the same lines as S77 and another where we include the R(2) line in each band. Table~\ref{gara_summ} provides a summary of the various \gara measurements performed in this work, as well as that of S77. Our initial recreation of the \copr~\gara measurement agrees reasonably well with their result. While our \logt N(1) measurement is 0.13 dex larger than theirs, that difference is within their error bars. Including the R(2) lines in the fits again produces a lower N(1) value, in this case resulting in \pctc{}N(0) = 4.5 $\pm$ 17.0\% and N(1) = -17.5 $\pm$ 6.2\%. The column densities produced using the CHESS-4 data are all at least a 0.1 dex larger than the corresponding S77 values, but this difference is well within the error bars in each case.

\begin{table}
\caption{A Summary of \gara \mh Analyses
\label{gara_summ}}
\centering
\begin{tabular}{ c  c  c  c  c  c }
\hline
\rule{0pt}{2.5ex} Source & \logt N(\mh) & \logt N(0) & \logt N(1) & \logt N(2) & T$_{01}$ \\
		& [\logt~cm$^{-2}$] & [\logt~cm$^{-2}$] & [\logt~cm$^{-2}$] & [\logt~cm$^{-2}$] & [K] \\ \hline
\rule{0pt}{3ex} Savage+ 77 & 19.24 & 18.93 $\pm$ 0.23 & 18.94 $\pm$ 0.23 & -- & 79\\[1.1ex]
No R(2)$^{\rm a}$  & 19.31 $\pm$ 0.01 & 18.93 $\pm$ 0.06 & 19.07 $\pm$ 0.03 & -- & 92.3 $\pm$ 10.5\\[1.1ex]
w/ R(2)$^{\rm a}$  & 19.29 $\pm$ 0.02 & 18.95 $\pm$ 0.04 & 19.00 $\pm$ 0.01 & 17.6 $\pm$ 0.35 & 81.5 $\pm$ 3.4\\[1.1ex]
CHESS-4 to \jrot = 1 & 19.38\uderr{0.22}{0.18} & 19.03\uderr{0.16}{0.18} & 19.13\uderr{0.26}{0.19} & -- & 77 $\pm$ 33 \\[1.1ex]
CHESS-4 to \jrot = 2 & 19.36\uderr{0.14}{0.17} & 19.02\uderr{0.04}{0.16} & 19.08\uderr{0.19}{0.19} & 17.75\uderr{0.46}{0.23} & 81 $\pm$ 24 \\[1.1ex]
CHESS-4 to \jrot = 7 & 19.39\uderr{0.20}{0.19} & 19.03\uderr{0.16}{0.17} & 19.07\uderr{0.15}{0.18} & 17.69\uderr{0.41}{0.16} & 82 $\pm$ 9 \\ \hline
\multicolumn{6}{l}{{\bf Note:} Column density errors were calculated using \sigc.} \\
\multicolumn{6}{l}{$^{\rm a}$ Fits to the \copr data using the CHESS analysis code.} \\
\end{tabular}
\end{table}

\section{An Updated Average \Tk of Diffuse Clouds} \label{extended}
\subsection{The Extended data set}
While the \bsco analysis made a intriguing argument in favor of a systematic error in the S77 column densities, the results were less conclusive for \gara. The existence of such an offset could have an impact on the measured average \Tk of that data set, leading to a change in the value that has been the primary point of comparison for \mh-based measurements of the diffuse ISM for over four decades. To further explore this effect, we extend our analysis of sightlines beyond those observed by CHESS to better constrain the magnitude of the \pctc in the column density regime of the S77 data set. This could be done by running an analysis on the entire S77 catalog, similar to what was done for \bsco and \gara, but that process would be difficult given the quality of the data, particularly for objects with U2 observations, and the lack of information on the final background subtracted spectra that were used by S77. In addition to this, the \copr U1 data is, in general, limited to the (1-0) Lyman absorption band. While we can obtain a measurement of \pctc in this region by including the single R(2) line, the resulting constraint on N(1) is limited compared to the multiple \jrot = 1 and 2 lines available over a wider bandpass. 

To address these problems, we generate a data set comprised of observations from \copr and \fuse. We select the highest quality \copr objects from S77 for analysis. These objects are then used as a point of comparison to the selected \fuse objects, which sample a broader range in column density and have a larger bandpass. Assuming the two groups produce comparable \pctc measurements, we can combine them to generate a trend in \pctc as a function of N(\mh) that can be used to revise the diffuse ISM temperatures derived by S77.

For the extended \copr sample, objects were selected based off of their quoted error in N(0) and N(1), which we found acted as a proxy for the quality of the spectra. The final selection of objects had \logt N(\mh) ranging from 19.49--20.28 and were all observed using the U1 channel. Analysis of each sightline followed the same procedure as that of \bsco, described in \S\ref{bsco_comp}. Unlike \bsco, 8 of the 9 objects lacked a published background-corrected spectrum in the S77 paper. In those cases, we iterated on the placement of the continuum until we were able to achieve the same results as S77, within their quoted error. We again focused on recreating their measurements and not on an independent determination of the column densities. A summary of our modeled column densities for these objects can be seen in \Tref{allNhs}.

To create the \fuse sample, we used~\cite{Wakker06}, \cite{Gillmon06}, \cite{Rachford09}, and~\cite{Burgh10} to select a total of 13 objects with archival \fuse observations that span a range in \logt N(\mh) of 14.5--20.65. This upper limit agrees well with that of the S77 sample, which has a maximum \logt N(\mh) of 20.67. In all cases, objects were first selected for column density and then S/N. We attempted to only select sightlines that showed signs of a single \mh absorption velocity component. For the highest column density objects, we additionally selected for sightlines that had low CO/\mh ratios, as measured by~\cite{Burgh10}. This ensured that we were not creating a data set that contained both translucent and diffuse sightlines~\citep{Snow06}, minimizing the effects of potential differences in \Tk between cloud populations.

The pre-processing of the \fuse objects roughly followed that of~\cite{Wakker06}. All observations, with the exception of HD 186994, used the LWRS channel. For HD 186994, data from the HIRS channel was used. All available observations for a given object were first co-added by channel and then binned by 3--4 pixels to avoid oversampling the data. This resulted in $\sim$0.04 \AA~wide pixels, which is about 1 bin per resolution element. The two LiF channels alone provided enough wavelength coverage that we chose to only use those for our analysis. A continuum for each spectrum was constructed by combining at least five splines, ranging in length from 10--40 \AA, with the length of an individual section being determined by the variability in the slope of the continuum. In rare cases where the shape of the continuum was masked by a large absorption features, a polynomial fit was used instead. Peak fluxes above the absorption features were estimated in an effort to correctly level the continuum across the gap.  A summary of our modeled column densities for these objects can be seen in \Tref{allNhs}.

Once normalized, the spectra were fit using the CHESS analysis code and an R = 15,000 Gaussian kernel~\citep{Wakker06} for the instrument profile. We found that our fit results were initially being skewed by the comparatively small error bars in the troughs of the absorption features. To remedy this, we set the values of any error bar smaller than the average error equal to the average standard deviation of a section of unabsorbed continuum. Like in the procedure used for the CHESS observations, non-\mh absorption features were masked. This occasionally resulted in the masking of one or more low \jrot lines of interest, but all objects that were used in our final analysis had at least three bands of R(0), R(1), and P(1) absorption features included in the fitting routine.

\subsection{Percent Change Measurements}
\begin{figure}
\centering
\includegraphics[width=0.6\textwidth]{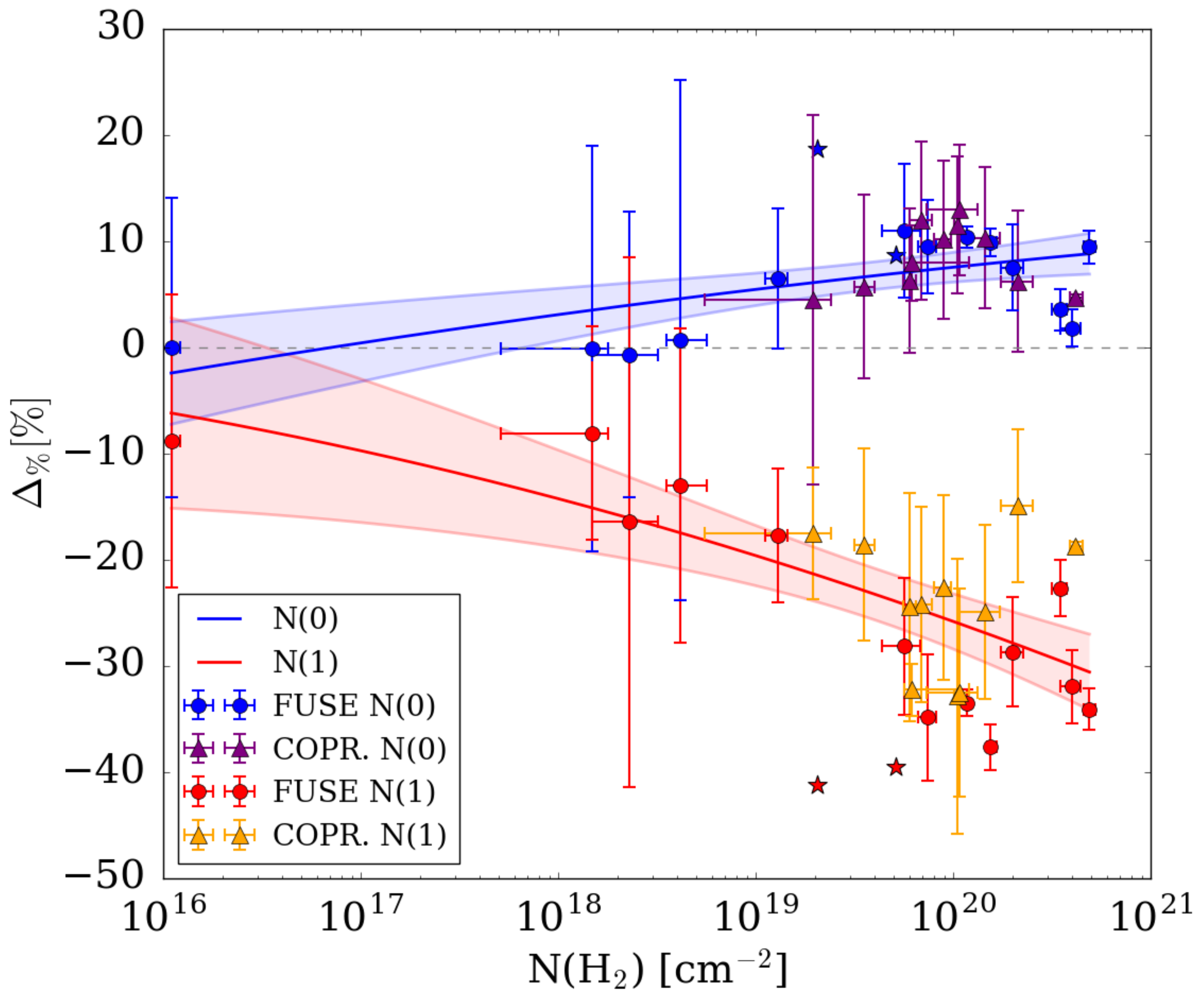}
\caption{The \pctc{}N(0) (blue circles for \fuse, purple triangles for \copr) and N(1) (red circles for \fuse, orange triangles for \copr) when the \jrot = 2 level is included in the model, as a function of total \mh column density. Second order polynomials have been fit to each distribution and are plotted, along with the surrounding 95\% confidence intervals, as a blue line for N(0) and a red line for N(1). The \pctc as measured using the CHESS observations are plotted as blue and red stars. They are not included in the polynomial fits.} 
\label{dNpct}
\end{figure}

The resulting \pctc{}N(0) and N(1) for the extended data set are listed in \Tref{pctch} and shown in \Fref{dNpct}. In all cases, the quoted values are comparing the fit results when \jrot = 2 is and is not included. We see that the measured \pctc agree well between the \copr and \fuse objects. We also find the expected trend of \pctc increasing in magnitude with column density, with the value becoming more positive for N(0) and more negative for N(1). For N(\mh) $\lesssim$ 10$^{18}$ cm$^{-2}$, N(0) does not appear to be greatly impacted by the inclusion of \jrot = 2. N(1) maintains a \pctc $\sim$ 10\% down to the lowest column densities measured in this work. While in both cases the error bars in this N(\mh) region are large, we caution that their values are likely overestimated (\S\ref{bsco_comp}) and note that their \pctc values agree with expectations. Mainly, the P(1) and R(2) lines are in close proximity to one another, particularly in the low vibrational bands. For example, the two lines are 0.06 \AA~apart in the (0-0) band, while the FWHM of their individual absorption features are on the order of 0.1 \AA~for the low N(\mh) objects. This is compared to the 0.5 \AA~separation between R(0) and R(1) in the same band. This means that N(0) is able to decouple from the relationship between N(1) and N(2) at low total column densities, allowing the \pctc{}N(0) to decay to zero while N(1) continues to be impacted by inclusion of \jrot = 2.

\begin{table}
\movetabledown=2.0in
\small
\begin{rotatetable}
\begin{center}
\caption{Measured N(\mh) Using the CHESS Analysis Pipeline
\label{allNhs}}
\begin{tabular}{ l | c  c  c | c  c  c  c  c  c }
\hline
\rule{0pt}{2.5ex}Name & \logt N(\mh) I & \logt N(\mh) II & Source$^{\rm a}$ & \logt N(\mh) & \logt N(0) & \logt N(1) & \logt N(2) & b$^{\rm b}$ & \Tk \\
		 & [\logt~cm$^{-2}$] & [\logt~cm$^{-2}$] &  & [\logt~cm$^{-2}$] & [\logt~cm$^{-2}$] & [\logt~cm$^{-2}$] & [\logt~cm$^{-2}$] & [km s$^{-1}$] & [K] \\ \hline
\multicolumn{10}{c}{\copr} \\ \hline
\multicolumn{1}{c}{} & \multicolumn{3}{| c |}{Previous Works} & \multicolumn{6}{c}{This Work} \\ \hline
\rule{0pt}{2.5ex} HD 149757 & 20.65 $\pm$ 0.08 & -- & 1 & 20.62\uderr{0.04}{0.03} & 20.49\uderr{0.04}{0.03} & 20.01\uderr{0.02}{0.02} & 18.41\uderr{0.03}{0.03} & $\leq$2.0 & 52 $\pm$ 1\\
\rule{0pt}{2.5ex} HD 167264 & 20.28 $\pm$ 0.10 & -- & 1 & 20.31\uderr{0.09}{0.08} & 20.02\uderr{0.09}{0.08} & 19.99\uderr{0.09}{0.07} & 18.05\uderr{1.67}{0.44} & 5.5 $\pm$ 3.2 & 75 $\pm$ 1 \\ 
\rule{0pt}{2.5ex} HD 112244 & 20.14 $\pm$ 0.11 & -- & 1 & 20.12\uderr{0.10}{0.11} & 19.85\uderr{0.09}{0.10} & 19.79\uderr{0.11}{0.08} & 17.95\uderr{0.66}{0.87} & 8.6 $\pm$ 6.6 & 73 $\pm$ 1 \\ 
\rule{0pt}{2.5ex} HD 144470 & 20.05 $\pm$ 0.11 & -- & 1 & 19.98\uderr{0.02}{0.02} & 19.80\uderr{0.01}{0.01} & 19.50\uderr{0.03}{0.03} & 16.09\uderr{0.18}{0.18} & 10.0$\pm$ 0.2 & 59 $\pm$ 1 \\ 
\rule{0pt}{2.5ex} HD 188209 & 20.01 $\pm$ 0.11 & -- & 1 & 19.99\uderr{0.13}{0.13} & 19.77\uderr{0.12}{0.14} & 19.57\uderr{0.12}{0.08} & 18.31\uderr{1.04}{0.44} & 6.1 $\pm$ 4.1 & 64 $\pm$ 3 \\ 
\rule{0pt}{2.5ex} HD 145502 & 19.89 $\pm$ 0.15 & -- & 1 & 19.91\uderr{0.05}{0.05} & 19.60\uderr{0.03}{0.03} & 19.61\uderr{0.07}{0.06} & 17.61\uderr{0.59}{0.44} & 4.8 $\pm$ 1.1 & 78 $\pm$ 3 \\ 
\rule{0pt}{2.5ex} HD 113904B & 19.83 $\pm$ 0.11 & -- & 1 & 19.79\uderr{0.07}{0.06} & 19.44\uderr{0.05}{0.04} & 19.53\uderr{0.09}{0.07} & 16.13\uderr{0.17}{0.38} & 9.0 $\pm$ 1.5 & 86 $\pm$ 4 \\ 
\rule{0pt}{2.5ex} HD 135591 & 19.77 $\pm$ 0.11 & -- & 1 & 19.75\uderr{0.03}{0.03} & 19.54\uderr{0.02}{0.02} & 19.33\uderr{0.05}{0.04} & 15.89\uderr{0.13}{0.27} & 7.1 $\pm$ 1.1 & 64 $\pm$ 2 \\ 
\rule{0pt}{2.5ex} HD 164402 & 19.49 $\pm$ 0.18 & -- & 1 & 19.51\uderr{0.06}{0.05} & 19.11\uderr{0.05}{0.04} & 19.28\uderr{0.07}{0.04} & 17.11\uderr{1.13}{0.63} & 3.7 $\pm$ 1.0 & 94 $\pm$ 3 \\ \hline
\multicolumn{10}{c}{\fuse} \\ \hline
\rule{0pt}{2.5ex} HD 157857 & 20.69 $\pm$ 0.09 & -- & 2 & 20.69\uderr{0.03}{0.03} & 20.33\uderr{0.03}{0.03} & 20.42\uderr{0.03}{0.03} & 19.02\uderr{0.02}{0.03} & $\leq$2.0 & 85 $\pm$ 1 \\
\rule{0pt}{2.5ex} HD 102065 & 20.53 $\pm$ 0.10 & 20.50 $\pm$ 0.06 & 2,3 & 20.54\uderr{0.04}{0.04} & 20.28\uderr{0.04}{0.04} & 20.18\uderr{0.04}{0.04} & 18.64\uderr{0.05}{0.06} & 3.0 $\pm$ 0.1 & 70 $\pm$ 1 \\
\rule{0pt}{2.5ex} HD 152590 & 20.51 $\pm$ 0.09 & -- & 2 & 20.60\uderr{0.06}{0.06} & 20.42\uderr{0.07}{0.07} & 20.11\uderr{0.03}{0.03} & 18.69\uderr{0.06}{0.06} & 3.8 $\pm$ 0.2 & 59 $\pm$ 2 \\
\rule{0pt}{2.5ex} HD 99857 & 20.25 $\pm$ 0.10  & -- & 2 & 20.30\uderr{0.06}{0.06} & 19.99\uderr{0.08}{0.07} & 19.99\uderr{0.04}{0.05} & 18.49\uderr{0.14}{0.05} & 4.9 $\pm$ 1.8 & 78 $\pm$ 3  \\
\rule{0pt}{2.5ex} HD 218915 & 20.16 $\pm$ 0.10 & -- & 2 & 20.19\uderr{0.03}{0.03} & 19.92\uderr{0.03}{0.03} & 19.81\uderr{0.01}{0.01} & 18.64\uderr{0.03}{0.04} & 3.6 $\pm$ 0.2 & 70 $\pm$ 1 \\
\rule{0pt}{2.5ex} HD 104705 & 20.00 $\pm$ 0.10 & -- & 2 & 20.07\uderr{0.01}{0.01} & 19.72\uderr{0.02}{0.02} & 19.77\uderr{0.01}{0.01} & 18.58\uderr{0.02}{0.02} & 4.0 $\pm$ 0.1 & 81 $\pm$ 1 \\
\rule{0pt}{2.5ex} NGC 7469 & 19.76\uderr{0.05}{0.04} & 19.67\uderr{0.10}{0.10} & 4,5 & 19.87\uderr{0.05}{0.05} & 19.60\uderr{0.05}{0.05} & 19.51\uderr{0.04}{0.05} & 18.33\uderr{0.06}{0.07} & 2.5 $\pm$ 0.2 & 71 $\pm$ 1 \\
\rule{0pt}{2.5ex} HD 186994 & 19.59 $\pm$ 0.04 & -- &  3 & 19.75\uderr{0.10}{0.11} & 19.32\uderr{0.11}{0.11} & 19.50\uderr{0.09}{0.09} & 18.13\uderr{0.21}{0.38} & 5.0 $\pm$ 0.3 & 96 $\pm$ 2 \\
\rule{0pt}{2.5ex} Mrk 335 & 19.07\uderr{0.07}{0.07} & 18.83\uderr{0.80}{0.80} & 4, 5 & 19.11\uderr{0.06}{0.06} & 18.70\uderr{0.05}{0.05} & 18.90\uderr{0.07}{0.07} & 16.24\uderr{0.03}{0.04} & 5.4 $\pm$ 0.1 & 99 $\pm$ 2 \\
\rule{0pt}{2.5ex} PG 0844+349 & 18.56\uderr{0.09}{0.09} & 18.22\uderr{0.18}{0.28} & 4,5 & 18.62\uderr{0.19}{0.07} & 18.04\uderr{0.17}{0.02} & 18.49\uderr{0.13}{0.09} & 15.74\uderr{1.98}{0.21} & 4.5 $\pm$ 2.5 & 148 $\pm$ 20 \\
\rule{0pt}{2.5ex} NGC 1068$^{\rm c}$ & 18.07\uderr{0.30}{0.43} & 18.13\uderr{0.13}{0.17} & 4,5 & 18.36\uderr{0.22}{0.19} & 18.08\uderr{0.09}{0.12} & 18.02\uderr{0.21}{0.29} & 16.39\uderr{1.13}{0.36} & 3.0 $\pm$ 1.0 & 73 $\pm$ 11 \\
\rule{0pt}{2.5ex} NGC 4151 & 16.60\uderr{0.54}{0.16} & 16.70\uderr{0.93}{0.31} & 4,5 & 18.17\uderr{0.10}{0.46} & 17.48\uderr{0.12}{0.50} & 18.07\uderr{0.09}{0.45} & 16.36\uderr{0.12}{0.52} & 4.0 $\pm$ 0.1 & 203 $\pm$ 27 \\
\rule{0pt}{2.5ex} PKS 0405-12 & 16.01\uderr{0.28}{0.14} & 15.44\uderr{0.18}{0.12} & 4,5 & 16.04\uderr{0.05}{0.01} & 15.24\uderr{0.05}{0.01} & 15.84\uderr{0.04}{0.03} & 15.19\uderr{0.07}{0.02} & 7.2 $\pm$ 0.9 & 208 $\pm$ 28 \\ \hline
\multicolumn{10}{p{0.95\textheight}}{$^{\rm a}$ Multiple sources are listed for objects with more than one independently measured column density. The sources are: (1)~\cite{Savage77}; (2)~\cite{Burgh10}; (3)~\cite{Rachford09}; (4)~\cite{Wakker06}; (5)~\cite{Gillmon06}. } \\
\multicolumn{10}{l}{$^{\rm b}$ Fits that returned the lowest possible b = 2 km s$^{-1}$ are quoted as upper limits.} \\
\multicolumn{10}{l}{$^{\rm c}$ Fit using an R = 6,600 Gaussian kernel, following~\cite{Wakker06}.}
\end{tabular}
\end{center}
\end{rotatetable}
\end{table}

To quantify the evolution of \pctc with N(\mh) we fit the two trends using second-order polynomials in \logt N(\mh) space. The resulting curves are plotted in \Fref{dNpct} along with the corresponding 95\% confidence intervals on the predicted values. The equations for these curves were found to be:
\begin{equation}
\Delta_{\%} {\rm N}(0) = -0.145 (\logtm {\rm N}(0))^{2} + 7.750 \logtm {\rm N}(0) - 89.358
\end{equation}
\begin{equation}
\Delta_{\%} {\rm N}(1) = -0.420 (\logtm {\rm N}(1))^{2} + 10.187 \logtm {\rm N}(1) - 61.475
\end{equation}
While the accuracy of our trends may not be high enough to provide meaningful updates on the level of an individual object, the corrections should provide a good estimate to the average properties of the S77 sample, mainly \Tk. When applying our fits to the N(0) and N(1) values of S77, we calculate a new \Tk = 68 $\pm$ 13 K. This value is 9 K lower than that of S77, but still within the error bars of both measurements. This updated value is in strong agreement with the~\cite{Rachford02,Rachford09} \fuse observations, who measured values of 67 and 68 K, respectively, and consistent with the values measured by other works such as~\cite{Burgh07} (\Tk = 74 $\pm$ 24 K) and~\cite{Sheffer08} (\Tk = 76 $\pm$ 17 K).

With our updates to N(0) and N(1), it is further useful to explore the impact of the \jrot = 2 level on N(\mh) since those values are important for abundance and extinction studies (see, e.g.~\citealt{Smith91,Indriolo13,Snow80,Kainulainen13}). In \Tref{pctch}, we see that our \jrot $\leq$ 2 models generally produce N(\mh) values that are on the order of 5\% smaller than those of S77, although this is not true in all cases. Comparing those two measurements does not provide a perfect correction because we were typically not able to exactly reproduce the S77 column densities in the \jrot $\leq$ 1 case and so a direct comparison may not fully reflect the changes that occur once \jrot = 2 is introduced.

\begin{figure}
\centering
\includegraphics[width=0.6\textwidth]{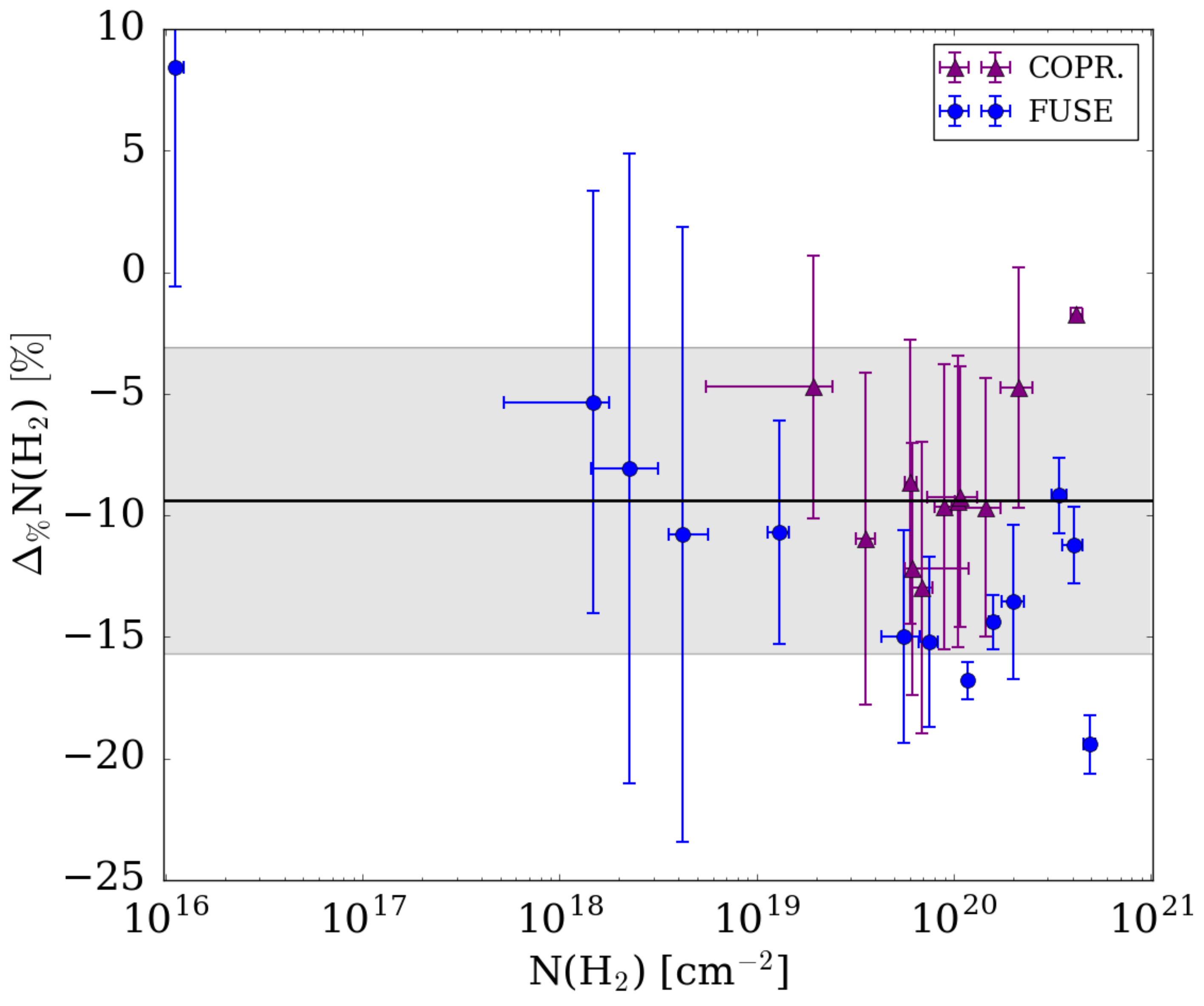}
\caption{The \pctc{}N(\mh) (blue circles for \fuse, purple triangles for \copr) when the \jrot = 2 level is included in the model, as a function of total \mh column density. The black line shows the average \pctc{} of -9.4\% and the gray shaded region extends over the $\pm$1$\sigma$ = 6.3\% range. These values were calculated using sightlines with N(\mh) $>$ 10$^{17}$ cm$^{-2}$.}
\label{dNtot}
\end{figure}

To better estimate how N(\mh) is impacted by the inclusion of \jrot = 2, we perform the same \pctc{} analysis described above. The results are shown in \Fref{dNtot}. We find that all of the sightlines, except for PKS 0405-12 (N(\mh) = 10$^{16.01}$ cm$^{-2}$), have a negative percent change. For PKS 0405-12, the model that includes \jrot = 2 produces an N(2) value that is non-negligible compared to N(0) and N(1), resulting in a significant increase N(\mh) despite the corresponding decrease in N(1). This discrepancy may be more indicative of the minimum detectable column density within the noise floor and less of an actual trend for low column density sightlines. For the remaining sightlines, \pctc{}N(\mh) does not vary significantly as a function of N(\mh). Instead, they are distributed around an average value of \pctc{}N(\mh) = -9.4 $\pm$ 6.3\%, indicating the S77 models likely produced N(\mh) values that were systematically too large by 5--10\%.

\begin{table}
\caption{Percent Change in N(0) and N(1) with the Inclusion of \jrot = 2
\label{pctch}}
\footnotesize
\centering
\begin{tabular}{|l|ccc|ccc|}
\hline
\rule{0pt}{2.5ex}Name & \logt N(0) & \logt N(0) & \pctc{}N(0) & \logt N(1) & \logt N(1) & \pctc{}N(1) \\
		 & (J$^{\prime \prime}_{max}$ = 1) & (J$^{\prime \prime}_{max}$ = 2) & & (J$^{\prime \prime}_{max}$ = 1) & (J$^{\prime \prime}_{max}$ = 2) & \\
		 & [\logt~cm$^{-2}$] & [\logt~cm$^{-2}$] & [\%] & [\logt~cm$^{-2}$] & [\logt~cm$^{-2}$] & [\%] \\ \hline
\multicolumn{7}{| c |}{\copr} \\ \hline
\rule{0pt}{2.5ex}HD 149757&20.47 $\pm$ $<$0.01& 20.49 $\pm$ $<$0.01 & 4.7 $\pm$ 0.3 & 20.10 $\pm$ $<$0.01 & 20.01 $\pm$ $<$0.01 & -18.7 $\pm$ 0.4\\
\rule{0pt}{2.5ex}HD 167264&20.00 $\pm$ 0.02& 20.02 $\pm$ 0.02 & 6.2 $\pm$ 6.7 & 20.06 $\pm$ 0.02 & 19.99 $\pm$ 0.02 & -14.9 $\pm$ 7.2\\ 
\rule{0pt}{2.5ex}HD 112244&19.80 $\pm$ 0.02& 19.85 $\pm$ 0.02 & 10.3 $\pm$ 6.6 & 19.91 $\pm$ 0.02 & 19.79 $\pm$ 0.02 & -24.9 $\pm$ 8.2\\ 
\rule{0pt}{2.5ex}HD 144470&19.75 $\pm$ 0.02& 19.80 $\pm$ 0.02 & 11.5 $\pm$ 6.4 & 19.67 $\pm$ 0.02 & 19.50 $\pm$ 0.03 & -32.8 $\pm$ 13.0\\ 
\rule{0pt}{2.5ex}HD 188209&19.71 $\pm$ 0.02& 19.77 $\pm$ 0.02 & 13.0 $\pm$ 6.2 & 19.74 $\pm$ 0.02 & 19.57 $\pm$ 0.02 & -32.5 $\pm$ 9.8\\ 
\rule{0pt}{2.5ex}HD 145502&19.56 $\pm$ 0.03& 19.60 $\pm$ 0.02 & 10.2 $\pm$ 7.4 & 19.72 $\pm$ 0.02 & 19.61 $\pm$ 0.02 & -22.6 $\pm$ 8.7\\ 
\rule{0pt}{2.5ex}HD 113904B&19.39 $\pm$ 0.03& 19.44 $\pm$ 0.02 & 12.0 $\pm$ 7.5 & 19.65 $\pm$ 0.02 & 19.53 $\pm$ 0.02 & -24.2 $\pm$ 9.2\\ 
\rule{0pt}{2.5ex}HD 135591&19.51 $\pm$ 0.02& 19.54 $\pm$ 0.02 & 6.3 $\pm$ 6.8 & 19.45 $\pm$ 0.02 & 19.33 $\pm$ 0.03 & -24.4 $\pm$ 10.8\\ 
\rule{0pt}{2.5ex}HD 164402&19.09 $\pm$ 0.03& 19.11 $\pm$ 0.03 & 5.7 $\pm$ 8.6 & 19.37 $\pm$ 0.02 & 19.28 $\pm$ 0.02 & -18.6 $\pm$ 9.1\\ 
\rule{0pt}{2.5ex}\bsco&19.48 $\pm$ 0.01& 19.52 $\pm$ 0.01 & 8.0 $\pm$ 3.6 & 19.59 $\pm$ 0.01 & 19.47 $\pm$ 0.01 & -32.3 $\pm$ 2.5\\
\rule{0pt}{2.5ex}\gara&18.93 $\pm$ 0.06& 18.95 $\pm$ 0.04 & 4.5 $\pm$ 17.4 & 19.07 $\pm$ 0.03 & 19.00 $\pm$ 0.01 & -17.5 $\pm$ 6.2\\ \hline
\multicolumn{7}{| c |}{\fuse} \\ \hline
\rule{0pt}{2.5ex}HD 157857&20.34 $\pm$ 0.01& 20.38 $\pm$ $<$0.01 & 9.5 $\pm$ 1.6 & 20.62 $\pm$ $<$0.01 & 20.44 $\pm$ $<$0.01 & -34.1 $\pm$ 2.0\\
\rule{0pt}{2.5ex}HD 102065&20.28 $\pm$ 0.01& 20.30 $\pm$ 0.01 & 3.6 $\pm$ 1.9 & 20.30 $\pm$ 0.01 & 20.19 $\pm$ 0.01 & -22.7 $\pm$ 2.6\\
\rule{0pt}{2.5ex}HD 152590&20.48 $\pm$ 0.01& 20.49 $\pm$ 0.01 & 1.8 $\pm$ 1.8 & 20.29 $\pm$ 0.01 & 20.13 $\pm$ 0.01 & -31.9 $\pm$ 3.4\\
\rule{0pt}{2.5ex}HD 99857&19.99 $\pm$ 0.01& 20.02 $\pm$ 0.01 & 7.5 $\pm$ 4.0 & 20.15 $\pm$ 0.01 & 20.00 $\pm$ 0.01 & -28.7 $\pm$ 5.1\\
\rule{0pt}{2.5ex}HD 218915&19.95 $\pm$ $<$0.01& 19.99 $\pm$ $<$0.01 & 9.9 $\pm$ 1.3 & 20.02 $\pm$ $<$0.01 & 19.82 $\pm$ $<$0.01 & -37.6 $\pm$ 2.2\\
\rule{0pt}{2.5ex}HD 104705&19.71 $\pm$ $<$0.01& 19.75 $\pm$ $<$0.01 & 10.4 $\pm$ 1.0 & 19.95 $\pm$ $<$0.01 & 19.77 $\pm$ $<$0.01 & -33.5 $\pm$ 1.2\\
\rule{0pt}{2.5ex}NGC 7469&19.57 $\pm$ 0.02& 19.61 $\pm$ 0.01 & 9.5 $\pm$ 4.4 & 19.70 $\pm$ 0.01 & 19.51 $\pm$ 0.01 & -34.9 $\pm$ 5.9\\
\rule{0pt}{2.5ex}HD 186994&19.33 $\pm$ 0.02& 19.37 $\pm$ 0.02 & 10.3 $\pm$ 6.4 & 19.67 $\pm$ 0.01 & 19.52 $\pm$ 0.02 & -29.7 $\pm$ 6.7\\
\rule{0pt}{2.5ex}Mrk 335&18.68 $\pm$ 0.02& 18.70 $\pm$ 0.02 & 6.5 $\pm$ 6.6 & 18.99 $\pm$ 0.02 & 18.91 $\pm$ 0.02 & -17.7 $\pm$ 6.3\\
\rule{0pt}{2.5ex}PG 0844+349&18.04 $\pm$ 0.08& 18.04 $\pm$ 0.07 & 0.7 $\pm$ 24.5 & 18.56 $\pm$ 0.04 & 18.5 $\pm$ 0.04 & -13.0 $\pm$ 14.8\\
\rule{0pt}{2.5ex}NGC 1068&18.09 $\pm$ 0.04& 18.09 $\pm$ 0.04 & -0.7 $\pm$ 13.4 & 18.07 $\pm$ 0.06 & 17.99 $\pm$ 0.07 & -16.4 $\pm$ 24.9\\
\rule{0pt}{2.5ex}NGC 4151&17.48 $\pm$ 0.06& 17.48 $\pm$ 0.06 & -0.1 $\pm$ 19.1 & 18.11 $\pm$ 0.03 & 18.08 $\pm$ 0.03 & -8.1 $\pm$ 10.0\\
\rule{0pt}{2.5ex}PKS 0405-12&15.25 $\pm$ 0.04& 15.25 $\pm$ 0.04 & 0.0 $\pm$ 14.1 & 15.88 $\pm$ 0.04 & 15.84 $\pm$ 0.04 & -8.8 $\pm$ 13.8\\ \hline
\multicolumn{7}{l}{\bf Notes:} \\
\multicolumn{7}{l}{{\bf (1)} Quoted errors on column densities are \sigf.} \\
\multicolumn{7}{l}{{\bf (2)} Column densities here are rounded and so they may not exactly reproduce the quoted \pctc.} \\
\end{tabular}
\end{table}

\section{Summary} \label{conclusions}
FUV observations of \mh in diffuse clouds provide crucial details related to the on-going physical and chemical processes within them. The CHESS sounding rocket was designed to be a pathfinder instrument for future FUV echelle spectrographs, leveraging a high resolution and large bandpass to observe individual diffuse and translucent clouds along sightlines towards bright stars.

Despite grating fabrication errors, CHESS-3 and CHESS-4 fulfilled the basic science goals of the instrument. Specifically, The two launches provided updated measurements of the \mh along the sightlines towards objects that had not been observed at $\lambda <$ 1150 \AA~since \copr. The inclusion of a curved echelle on CHESS-4 (the first such application in space-borne astronomy) additionally helped mitigate the resolution issue. Continued development of such a concept could improve spectrograph performance since the extra degrees of freedom could allow for further aberration control. 

For CHESS-3, we found N(\mh) and N(0) agreed well with that of S77, but our N(1) results differed by $\sim$0.4 dex. This discrepancy lead to a reevaluation of the S77 results. In particular, the exclusion of the (1-0)R(2) line in their analysis likely caused a larger inferred N(1) value. This conclusion was supported by our CHESS-4 observations (although, with a larger uncertainty). To further explore this trend, we generated an extended data set comprised of \fuse and \copr observations and modeled their \mh column densities following our CHESS procedure. We found that N(0) and N(1) were both impacted by the inclusion of the \jrot = 2 level in the model, particularly for N(\mh) $\gtrsim$ 10$^{18.0}$ cm$^{2}$. The magnitude of this effect further scaled with N(\mh) (\Fref{dNpct}). By applying our measured trend to the values produced by S77, we find an updated average \Tk of 68 $\pm$ 13 K for their sample of diffuse sightlines. In light of these results, we caution against imposing limits on \jrot when modeling \mh absorption lines since the interdependence of the \jrot levels can significantly impact the resulting column densities.

\section*{Acknowledgments}       
 
The authors would like to thank the students and staff at CU for their tremendous help in seeing CHESS-3 and CHESS-4 come to fruition. We would also like to thank the NSROC staff at WFF, WSMR, and on Roi-Namur for their tireless efforts that pushed us to two smooth launches. This work was supported by NASA grants NNX13AF55G and NNX16AG28G to the University of Colorado. Some of the data presented in this paper were obtained from the Mikulski Archive for Space Telescopes (MAST). STScI is operated by the Association of Universities for Research in Astronomy, Inc., under NASA contract NAS5-26555. Support for MAST is provided by the NASA Office of Space Science via grant NNX13AC07G and by other grants and contracts. 

\newpage

\bibliography{references}

\begin{thebibliography}{}
\expandafter\ifx\csname natexlab\endcsname\relax\def\natexlab#1{#1}\fi
\providecommand{\url}[1]{\href{#1}{#1}}
\providecommand{\dodoi}[1]{doi:~\href{http://doi.org/#1}{\nolinkurl{#1}}}
\providecommand{\doeprint}[1]{\href{http://ascl.net/#1}{\nolinkurl{http://ascl.net/#1}}}
\providecommand{\doarXiv}[1]{\href{https://arxiv.org/abs/#1}{\nolinkurl{https://arxiv.org/abs/#1}}}

\bibitem[{{Abt}(1981)}]{Abt81}
{Abt}, H.~A. 1981, \apjs, 45, 437, \dodoi{10.1086/190719}

\bibitem[{{Beasley} {et~al.}(2010){Beasley}, {Burgh}, \& {France}}]{Beasley10}
{Beasley}, M., {Burgh}, E., \& {France}, K. 2010, in \procspie, Vol. 7732,
  Space Telescopes and Instrumentation 2010: Ultraviolet to Gamma Ray, 773206

\bibitem[{{Bohlin} {et~al.}(1983){Bohlin}, {Jenkins}, {Spitzer}, {York},
  {Hill}, {Savage}, \& {Snow}}]{Bohlin83}
{Bohlin}, R.~C., {Jenkins}, E.~B., {Spitzer}, Jr., L., {et~al.} 1983, \apjs,
  51, 277

\bibitem[{{Burgh} {et~al.}(2010){Burgh}, {France}, \& {Jenkins}}]{Burgh10}
{Burgh}, E.~B., {France}, K., \& {Jenkins}, E.~B. 2010, \apj, 708, 334

\bibitem[{{Burgh} {et~al.}(2007){Burgh}, {France}, \& {McCandliss}}]{Burgh07}
{Burgh}, E.~B., {France}, K., \& {McCandliss}, S.~R. 2007, \apj

\bibitem[{{Drake} {et~al.}(1976){Drake}, {Jenkins}, {Bertaux}, {Festou}, \&
  {Keller}}]{Drake76}
{Drake}, J.~F., {Jenkins}, E.~B., {Bertaux}, J.~L., {Festou}, M., \& {Keller},
  H.~U. 1976, \apj, 209, 302, \dodoi{10.1086/154721}

\bibitem[{{Federman} {et~al.}(1980){Federman}, {Glassgold}, {Jenkins}, \&
  {Shaya}}]{Federman80}
{Federman}, S.~R., {Glassgold}, A.~E., {Jenkins}, E.~B., \& {Shaya}, E.~J.
  1980, \apj, 242, 545

\bibitem[{{Ferri{\`e}re}(2001)}]{Ferriere01}
{Ferri{\`e}re}, K.~M. 2001, Reviews of Modern Physics, 73, 1031,
  \dodoi{10.1103/RevModPhys.73.1031}

\bibitem[{{France} {et~al.}(2016){France}, {Hoadley}, {Fleming}, {Kane},
  {Nell}, {Beasley}, \& {Green}}]{France16a}
{France}, K., {Hoadley}, K., {Fleming}, B.~T., {et~al.} 2016, Journal of
  Astronomical Instrumentation, 5, 1640001, \dodoi{10.1142/S2251171716400018}

\bibitem[{{France} {et~al.}(2013){France}, {Nell}, {Kane}, {Burgh}, {Beasley},
  \& {Green}}]{France13a}
{France}, K., {Nell}, N., {Kane}, R., {et~al.} 2013, \apjl, 772, L9

\bibitem[{{Gillmon} {et~al.}(2006){Gillmon}, {Shull}, {Tumlinson}, \&
  {Danforth}}]{Gillmon06}
{Gillmon}, K., {Shull}, J.~M., {Tumlinson}, J., \& {Danforth}, C. 2006, \apj,
  636, 891, \dodoi{10.1086/498053}

\bibitem[{{Hoadley} {et~al.}(2019){Hoadley}, {France}, {Nell}, {Kane},
  {Fleming}, \& {Youngblood}}]{Hoadley19}
{Hoadley}, K., {France}, K., {Nell}, N., {et~al.} 2019, Experimental Astronomy

\bibitem[{{Hoadley} {et~al.}(2014){Hoadley}, {France}, {Nell}, {Kane},
  {Schultz}, {Beasley}, {Green}, {Kulow}, {Kersgaard}, \&
  {Fleming}}]{Hoadley14}
{Hoadley}, K., {France}, K., {Nell}, N., {et~al.} 2014, in \procspie, Vol.
  9144, {Space Telescopes and Instrumentation 2014: Ultraviolet to Gamma Ray},
  914406

\bibitem[{{Hoadley} {et~al.}(2016){Hoadley}, {France}, {Kruczek}, {Fleming},
  {Nell}, {Kane}, {Swanson}, {Green}, {Erickson}, \& {Wilson}}]{Hoadley16}
{Hoadley}, K., {France}, K., {Kruczek}, N., {et~al.} 2016, in \procspie, Vol.
  9905, Space Telescopes and Instrumentation 2016: Ultraviolet to Gamma Ray,
  99052V

\bibitem[{{Indriolo} {et~al.}(2013){Indriolo}, {Neufeld}, {Seifahrt}, \&
  {Richter}}]{Indriolo13}
{Indriolo}, N., {Neufeld}, D.~A., {Seifahrt}, A., \& {Richter}, M.~J. 2013,
  \apj, 764, 188, \dodoi{10.1088/0004-637X/764/2/188}

\bibitem[{{Jenkins}(1975)}]{Jenkins75}
{Jenkins}, E.~B. 1975, Memo to Copernicus Astronomers and Guest Investigators:
  Computations of the U1 Instrument Profile

\bibitem[{{Jenkins} {et~al.}(1988){Jenkins}, {Joseph}, {Long}, {Zucchino}, \&
  {Carruthers}}]{Jenkins88}
{Jenkins}, E.~B., {Joseph}, C.~L., {Long}, D., {Zucchino}, P.~M., \&
  {Carruthers}, G.~R. 1988, in \procspie, Vol. 932, Ultraviolet technology II,
  ed. R.~E. {Huffman}, 213--229

\bibitem[{{Jenkins} \& {Peimbert}(1997)}]{Jenkins97}
{Jenkins}, E.~B., \& {Peimbert}, A. 1997, \apj, 477, 265,
  \dodoi{10.1086/303694}

\bibitem[{{Jenkins} {et~al.}(2000){Jenkins}, {Wo{\'z}niak}, {Sofia},
  {Sonneborn}, \& {Tripp}}]{Jenkins00}
{Jenkins}, E.~B., {Wo{\'z}niak}, P.~R., {Sofia}, U.~J., {Sonneborn}, G., \&
  {Tripp}, T.~M. 2000, \apj, 538, 275, \dodoi{10.1086/309130}

\bibitem[{{Kainulainen} \& {Tan}(2013)}]{Kainulainen13}
{Kainulainen}, J., \& {Tan}, J.~C. 2013, \aap, 549, A53,
  \dodoi{10.1051/0004-6361/201219526}

\bibitem[{{Kruczek} {et~al.}(2019){Kruczek}, {France}, {Fleming}, \&
  {Ulrich}}]{Kruczek19}
{Kruczek}, N., {France}, K., {Fleming}, B., \& {Ulrich}, S. 2019

\bibitem[{{Kruczek} {et~al.}(2017){Kruczek}, {Nell}, {France}, {Hoadley},
  {Fleming}, {Kane}, {Ulrich}, {Egan}, \& {Beatty}}]{Kruczek17}
{Kruczek}, N., {Nell}, N., {France}, K., {et~al.} 2017, in SPIE, Vol. 10397,
  Society of Photo-Optical Instrumentation Engineers (SPIE) Conference Series,
  103971G

\bibitem[{{Kruczek} {et~al.}(2018){Kruczek}, {Nell}, {France}, {Hoadley},
  {Fleming}, {Ulrich}, {Miller}, {Egan}, {Witt}, \& {Kane}}]{Kruczek18}
{Kruczek}, N., {Nell}, N., {France}, K., {et~al.} 2018, in SPIE, Vol. 10699,
  Space Telescopes and Instrumentation 2018: Ultraviolet to Gamma Ray, 106990K

\bibitem[{{Lacour} {et~al.}(2005){Lacour}, {Ziskin}, {H{\'e}brard}, {Oliveira},
  {Andr{\'e}}, {Ferlet}, \& {Vidal-Madjar}}]{Lacour05}
{Lacour}, S., {Ziskin}, V., {H{\'e}brard}, G., {et~al.} 2005, \apj, 627, 251,
  \dodoi{10.1086/430291}

\bibitem[{{McCandliss}(2003)}]{McCandliss03}
{McCandliss}, S.~R. 2003, \pasp, 115, 651, \dodoi{10.1086/375387}

\bibitem[{{Pellerin} {et~al.}(2002){Pellerin}, {Fullerton}, {Robert}, {Howk},
  {Hutchings}, {Walborn}, {Bianchi}, {Crowther}, \& {Sonneborn}}]{Pellerin02}
{Pellerin}, A., {Fullerton}, A.~W., {Robert}, C., {et~al.} 2002, \apjs, 143,
  159, \dodoi{10.1086/342268}

\bibitem[{{Prinja} {et~al.}(1997){Prinja}, {Massa}, {Fullerton}, {Howarth}, \&
  {Pontefract}}]{Prinja97}
{Prinja}, R.~K., {Massa}, D., {Fullerton}, A.~W., {Howarth}, I.~D., \&
  {Pontefract}, M. 1997, \aap, 318, 157

\bibitem[{{Rachford} {et~al.}(2002){Rachford}, {Snow}, {Tumlinson}, {Shull},
  {Blair}, {Ferlet}, {Friedman}, {Gry}, {Jenkins}, {Morton}, {Savage},
  {Sonnentrucker}, {Vidal-Madjar}, {Welty}, \& {York}}]{Rachford02}
{Rachford}, B.~L., {Snow}, T.~P., {Tumlinson}, J., {et~al.} 2002, \apj, 577,
  221, \dodoi{10.1086/342146}

\bibitem[{{Rachford} {et~al.}(2009){Rachford}, {Snow}, {Destree}, {Ross},
  {Ferlet}, {Friedman}, {Gry}, {Jenkins}, {Morton}, {Savage}, {Shull},
  {Sonnentrucker}, {Tumlinson}, {Vidal-Madjar}, {Welty}, \&
  {York}}]{Rachford09}
{Rachford}, B.~L., {Snow}, T.~P., {Destree}, J.~D., {et~al.} 2009, \apjs, 180,
  125, \dodoi{10.1088/0067-0049/180/1/125}

\bibitem[{{Rogerson} {et~al.}(1973){Rogerson}, {Spitzer}, {Drake}, {Dressler},
  {Jenkins}, {Morton}, \& {York}}]{Rogerson73}
{Rogerson}, J.~B., {Spitzer}, L., {Drake}, J.~F., {et~al.} 1973, \apjl, 181,
  L97, \dodoi{10.1086/181194}

\bibitem[{{Savage} {et~al.}(1977){Savage}, {Bohlin}, {Drake}, \&
  {Budich}}]{Savage77}
{Savage}, B.~D., {Bohlin}, R.~C., {Drake}, J.~F., \& {Budich}, W. 1977, \apj,
  216, 291, \dodoi{10.1086/155471}

\bibitem[{{Sheffer} {et~al.}(2008){Sheffer}, {Rogers}, {Federman}, {Abel},
  {Gredel}, {Lambert}, \& {Shaw}}]{Sheffer08}
{Sheffer}, Y., {Rogers}, M., {Federman}, S.~R., {et~al.} 2008, \apj, 687, 1075,
  \dodoi{10.1086/591484}

\bibitem[{{Siegmund} {et~al.}(2009){Siegmund}, {Tremsin}, \&
  {Vallerga}}]{Siegmund09}
{Siegmund}, O.~H.~W., {Tremsin}, A.~S., \& {Vallerga}, J.~V. 2009, in
  \procspie, Vol. 7435, UV, X-Ray, and Gamma-Ray Space Instrumentation for
  Astronomy XVI, 74350L

\bibitem[{{Smith} {et~al.}(1991){Smith}, {Bruhweiler}, {Lambert}, {Savage},
  {Cardelli}, {Ebbets}, {Lyu}, \& {Sheffer}}]{Smith91}
{Smith}, A.~M., {Bruhweiler}, F.~C., {Lambert}, D.~L., {et~al.} 1991, \apjl,
  377, L61, \dodoi{10.1086/186117}

\bibitem[{{Snow} \& {McCall}(2006)}]{Snow06}
{Snow}, T.~P., \& {McCall}, B.~J. 2006, \araa, 44, 367,
  \dodoi{10.1146/annurev.astro.43.072103.150624}

\bibitem[{{Snow} \& {Jenkins}(1980)}]{Snow80}
{Snow}, Jr., T.~P., \& {Jenkins}, E.~B. 1980, \apj, 241, 161,
  \dodoi{10.1086/158328}

\bibitem[{{Spitzer} \& {Cochran}(1973)}]{Spitzer73}
{Spitzer}, Jr., L., \& {Cochran}, W.~D. 1973, \apjl, 186, L23,
  \dodoi{10.1086/181349}

\bibitem[{{Thomas}(2003)}]{Thomas03}
{Thomas}, R.~J. 2003, in \procspie, Vol. 4853, Innovative Telescopes and
  Instrumentation for Solar Astrophysics, ed. S.~L. {Keil} \& S.~V. {Avakyan},
  411--418

\bibitem[{{Vallerga} {et~al.}(2010){Vallerga}, {Raffanti}, {Tremsin},
  {Siegmund}, {McPhate}, \& {Varner}}]{Vallerga10}
{Vallerga}, J., {Raffanti}, R., {Tremsin}, A., {et~al.} 2010, in \procspie,
  Vol. 7732, Space Telescopes and Instrumentation 2010: Ultraviolet to Gamma
  Ray, 773203

\bibitem[{{Wakker}(2006)}]{Wakker06}
{Wakker}, B.~P. 2006, \apjs, 163, 282, \dodoi{10.1086/500365}

\end{thebibliography}
\bibliographystyle{aasjournal}
\end{document}